\documentclass[aps,prd, nofootinbib,superscriptaddress, article,11pt]{revtex4}

\usepackage{amsmath, amsfonts, amsthm, amssymb, graphicx, slashed}
\usepackage{subfig}
\usepackage{epstopdf}

\def\be{\begin{equation}}
\def\ee{\end{equation}}
\def\ba{\begin{eqnarray}}
\def\ea{\end{eqnarray}}

\def\bdm{\begin{displaymath}}
\def\edm{\end{displaymath}}

\def\bq{\begin{quote}}
\def\eq{\end{quote}}

\def\ltap{\ \raise.3ex\hbox{$<$\kern-.75em\lower1ex\hbox{$\sim$}}\ }
\def\gtap{\ \raise.3ex\hbox{$>$\kern-.75em\lower1ex\hbox{$\sim$}}\ }
\def\gl{\ \raise.5ex\hbox{$>$}\kern-.8em\lower.5ex\hbox{$<$}\ }
\def\roughly#1{\raise.3ex\hbox{$#1$\kern-.75em\lower1ex\hbox{$\sim$}}}

 at 10truept

\newcommand{\beq}{\begin{equation}}
\newcommand{\eeq}{\end{equation}}
\newcommand{\bea}{\begin{eqnarray}}
\newcommand{\eea}{\end{eqnarray}}
\newcommand{\beqa}{\begin{eqnarray}}
\newcommand{\eeqa}{\end{eqnarray}}

\begin{document}
\title{  Holographic superconductors
with Weyl corrections}

\author         {Davood Momeni}\footnote{Corresponding author}
\email          {d.momeni@yahoo.com}
%\homepage       {http://inspirehep.net/author/profile/D.Momeni.1}
\affiliation    {Eurasian International Center for Theoretical Physics and Department of General \& Theoretical Physics, Eurasian National University, \\ Astana 010008, Kazakhstan}
\author        {Muhammad Raza}
\affiliation {Department of Mathematics, COMSATS Institute of Information Technology, Sahiwal, Pakistan
;\\ Centre for Optical and Electromagnetic Research, Department of Electrical Engineering, Zhejiang University, Hangzhou,
China }
\author         {Ratbay Myrzakulov}
\affiliation    {Eurasian International Center for Theoretical Physics and Department of General \& Theoretical Physics, Eurasian National University, \\ Astana 010008, Kazakhstan}
\date{\today}

\begin{abstract}
A quick review on the analytical aspects of holographic superconductors (HSC) with Weyl corrections has been presented. Mainly we focus on matching method and variations approaches. Different types of such HSC have been investigated, s-wave, p-wave and   St\'{u}ckelberg ones. We also review the fundamental construction of a p-wave type , in which the non-Abelian gauge field is coupled to the Weyl tensor. The results are compared from numerics to analytical results.
\end{abstract}

\maketitle
\tableofcontents

%%%%%%%%%%%%%%%%%%%%%%%%%
\section{Introduction}
%%%%%%%%%%%%%%%%%%%%%%%%%%%
Gauge/gravity duality or strong/weak duality is considered as a significant result of the anti-de
Sitter/conformal field theory correspondence (AdS/CFT)~\cite{Maldacena:1997re,Gubser:1998bc,Witten:1998qj}. Because this conjecture provides a bridge between two regimes of couplings, so it is very useful to investigate the basic properties of strongly coupled systems in condensed matter or plasma in $D$-dimension of studying the dual gravitational dual AdS space time in a higher dimension $ (D+1)$-dimension~\cite{Hartnoll:2009sz,Herzog:2009xv,McGreevy:2009xe,Horowitz:2010gk} for reviews). 
It is also possible to develop the scheme of AdS/CFT to the non relativistic regimes or systems with less numbers of symmetry ~\cite{Horowitz:2012ky,Horowitz:2013jaa,Ling:2013aya,
Donos:2011bh,Donos:2013wia,Rozali:2013ama,Cai:2013sua}. One good example is to investigate the non-equilibrium systems which have critically ~\cite{Murata:2010dx,Bhaseen:2012gg,Adams:2012pj,Garcia-Garcia:2013rha,Chesler:2013lia}.\par
The best example for us to see how this stringy inspired principle works in condensed matter are to study high temperature superconductors which have gravitational dual. Such superconductors are called as holographic superconductors \cite{Hartnoll:2008vx,Hartnoll:2008kx}. The hidden mechanism behind the superconductivity in the gauge / gravity picture is presented by the basic instability of AdS space time in the presence of the complex charged scalar field in bulk. The distorted black hole is a hairy black hole with scalar charge. In dual picture, the complex scalar hair corresponds to the condensation of the dual relevant operator $\mathcal{O}_{i}$. The reason is that the $U(1)$ symmetry is broken when the system enters the thermal phase with the temperatue $T<T_c$  where $T_c$ denotes the critical temperature of the system. The above description is for a s-wave type of HSC. The main idea is how to realize the phase transition from a normal phase (no hair black hole) to the hairy (conductor) phase. We can also generalize this fully consistent description to  the case with zero temperature,using AdS soliton~\cite{Nishioka:2009zj}. Here the  mass
gap  is presented to describe consistency a confining phase  which is happen in a dual
theory~\cite{Witten:1998zw}(see ~\cite{Horowitz:2010jq} for the complete description). \par
S-wave is not the only type of HSC. If one replaced $U (1) $ by $SU (2) $, the p-wave type appeared~\cite{Gubser:2008wv}. Instead of Maxwell's field $A_{\mu}$ into the bulk, we need to introduce a 
Yang-Mills field. The dual field of the vector order parameter $\hat{J}^{\mu}(x)$ is a gauge boson which is generated by a Higgs type mechanism (see ~\cite{Aprile:2010ge} for a realization of p-wave using the condensation of a tensor field). We can extend our studies by including the Backreaction effects. But the simplest case happens when we are working in the so called as the probe limit. By probe approximation we mean we tend the scalar charge $q\to \infty$ . But meanwhile we keep $qA_\mu$
finite and fixed.\par
Higher order curvature terms like Gauss-Bonnet (GB) $$G=R^{\alpha\beta\gamma\theta}R_{\alpha\beta\gamma\theta}
-4R^{\alpha\beta}R_{\alpha\beta}+R^2$$ or Weyl $$C_{\alpha\beta\mu\nu}=R_{\alpha\beta\mu\nu}+\frac{1}{2}\Big(-R_{\alpha\mu}g_{\beta\nu}+R_{\alpha\nu}g_{\beta\mu}+R_{\beta\mu}g_{\alpha\nu}-R_{\beta\nu}g_{\alpha\mu}+\frac{1}{3}(g_{\alpha\mu}g_{\beta\nu}-g_{\alpha\nu}g_{\beta\mu})R)\Big)$$ also provide a landscape for gravity part of AdS/CFT. Specially, GB corrected holographic superconductors has been investigated in several works \cite{Pan:2009xa}-\cite{Wu:2014lta}.
Weyl corrected HSC is the topic of our review. We will give a quick review on the subject. Our main focus here is on analytical properties of such superconductors.

%%%%%%%%%%%%%%%%%%%%%%%%%%%
\section{ A bird's-eye view  to $AdS/CFT$}
%%%%%%%%%%%%%%%%%%%%%%%%%%%
Our aim of this section is to develop the basic structure of applied AdS/CFT. Let us start with the simplest form of the Maldacena conjecture. As we understood, this conjecture learns us that \emph{the solutions of the type IIB superstring theory in $AdS_5\times S^5$  are related to 
the solutions of a super Yang-Mills (YM) theory} . This equivalence is a result of 
 \emph{large t'Hooft
coupling }limit in which  for the large numbers of colors $N$, the YM coupling and the color number is related to each other by 
 the fundamental equation $%
\lambda=g_{YM}^2 N$. From the geometrical point of view, 
we suppose that the manifold $AdS_5\times
S^5 $ is compact. From the dynamical point of view if we look at the action, 
we observe that the four (five) dimensional action is obtained 
by an  integration over a ten or eleven dimensional space time $%
AdS_5\times S^5$. 
% If somebody uses the noncompact manifold, consequently 
%the metric has an AdS boundary.
 What we learn from CFT is how to read the
expectation values of the dual operators $<\mathcal{O}_{i}>$ , living on the AdS
boundary. Precisely, there is an isomorphism between this set of CFT relevant operatoes and the classical bulk
fields, namely scalar field(s) $\psi$. As a handy tool, we are able to read
$<\mathcal{O}_{i}>$ from the asymptotic limit of the two point functions(Green's functions) of the bulk scalar field on $AdS_5$.
\par
To be more precise, \emph{there is an  isomorphism from some asymptotic values of the bulk field $\psi$ to the relevant
operator  $\mathcal{O}_{i}$}. So, since the energy momentum tensor of CFT is traceless, from holographic view, always it is possible  to find a unique metric 
 $g_{\mu \nu }$  in the bulk which it corresponds to that CFT's energy momentum tensor.  This later equivalence is called as 
\emph{ gauge/gravity duality}. 
Technically, if we look from dynamics, we always have a definite equivalence class of all possible \emph{on-shell} value ( the computed string action by substituting it by the classical solutions of field equations) of the total action of bulk fields and super string theory and dual action of relevant quantum CFT. To give an illustrative scheme on how this equivalence class appeared, let us to compute explicitly the relation between the two-point function of scalar bulk field and the expectation value of CFT operators. \par

Computation is stated by the equation of motion of the scalar field in the $AdS_ {5} $ in the following form
\begin{equation}
(\nabla _{\mu }\nabla ^{\mu }-m^{2})\psi =0.
\end{equation}%
The mass term $m^ {2} $ encodes the full history and information about the the $<\mathcal{O}_{i}>$. It is well known understood that on $AdS_ {5} $ background, the Green's function has the following simple form:
\begin{equation}
G(r;x,x^{\prime })=\frac{\Gamma (\Delta )}{\pi ^{2}\Gamma (\Delta -2)}(\frac{%
L^{2}}{r})^{\Delta }((\frac{L^{2}}{r})^{2}+|x-x^{\prime 2}|)^{-\Delta }.
\end{equation}%
Where $x^{\prime }$ is the retarded coordinate. Also the exponent $\Delta $ (is called as conformal dimension) is related to the mass  $m^{2}$ in 4D, through the following equation:
\begin{equation}
\Delta (\Delta -4)=m^{2}L^{2}.
\end{equation}%
By unitarity reason, we need to assume that 
$\Delta >2$ to keep  the scalar mass  in the BF bound. The asymptotic behavior of the 
 Green's function the vicinity of  the AdS boundary $r\rightarrow
\infty $ is given by :
\begin{equation}
G(r,x,x')\sim s(x)(rL^{-2})^{\Delta-4}+c(x)(rL^{-2})^{-\Delta}.
\end{equation}%
It is straightforward to show that in this asymptotic expression, the first function $s (x) $ plays the role of source in the scalar field $\psi $ when we evaluate this field on AdS (asymptotic) regime $r\to\infty$ and the second $c (x) $ defines$\mathcal{O}_{i}>$. 
It's the way and a passion to find quantum expectation values (on boundary) from classical solutions (in bulk).

%%%%%%%%%%%%%%%%%%%%%%%%%%
\section{Building s-wave holographic superconductor with Weyl corrections}
%%%%%%%%%%%%%%%%%%%%%%%%%%%
To build-up a model of s-wave the ingredients which we need is a set of $U (1) $ gauge field and a charged scalar field. We need the scalar field to provide dual quantum operators on CFT. Since the first non trivial effect of Weyl tensor is appeared in 5D, the holographic picture is described by $ (3+1) D$ dual theory. We assume that the bulk metric is asymptotically $AdS_5$ and for the simplicity we put the AdS radius $L=1$. Also since we will study the system in equilibrium and in the probe limit, to ignoring the Backreaction effects, we put the electric charge $q=1$. The simplest form of such systems with the mentioned assumptions is proposed as the following \cite{Wu:2010vr}:
\begin{eqnarray}
S=\int dtd^{4}x\sqrt{-g}\{\frac{1}{16\pi G_{5}}(R+12)-\frac{1}{4}%
(F^{\mu\nu}F_{\mu\nu}-4\gamma
C^{\mu\nu\rho\sigma}F_{\mu\nu}F_{\rho\sigma})-|D_{\mu}\psi|^2-m^2|\psi|^2\}\label{S1}
\end{eqnarray}
Here $G_{5}$ plays the role of an effective 5D  gravitational constant, next term beside $R$ is a constant which it measures the AdS radius,also the electromagnetic strength tensor is represented by  $F_{\mu\nu}=\partial_{\mu}A_{\nu}-\partial_{\nu}A_{%
\mu}$ where $A_{\mu}$ is the gauge field. Coordinates $x^{\mu}$ run from $0$ to $4$ are corresponding to the $t,r,x,y,z$. The model is similar to the GL but the square term of mass has different form with GL phenomenological Lagrangian. By the reason of a gauge field, we replace the covariant derivatives by the gauge derivative is defined by $D_{\mu}=\nabla_{\mu}-iA_{\mu}$. We should mention here the role of gauge field. Since this field lives at bulk, the expected role will be appeared at boundary through the relevant quantum operator. A vector field induces a vector current $J^{\mu}$ on the boundary. Such a conserved current has an expectation value $<J^{\mu}>$ which corresponds to the $U (1) $ gauge field.
This conserved dual current corresponds to the \emph{global U (1) symmetry}.
%%%%%%%%%%%%%%%%%%%%%%%%%%%%%
We must explain the role of the Weyl tensor in our model. Firstly we mention here that the classical solution of equations of motion of (\ref{S1}) is represented by the Einstein equation in the presence of the cosmological constant. As we know, there is only a unique type of tensor correction which can be coupled non minimally to the Weyl tensor. This coupling comes from the $F_{\mu\nu}$. The coupling term leads to the the operator of the dimension-six operator. To parameterize it we introduce a coupling constant $\gamma$. It is a remarkable comment saying it, other curvature coupling terms to the $F_{\mu\nu}$ are just a shift symmetry and so we can omit all safely. We are motivated by quantum corrections to have Weyl coupling. Actually thanks to the $1-$ loop quantum corrections, we know that 
In any smoothly Riemannian background when we integrate all types of (charged) matter fields with a suitable cutoff (threshold's mass), the Weyl coupling term $%
C^{\mu\nu\rho\sigma}F_{\mu\nu}F_{\rho\sigma}$ is provided at $1-$loop quantum corrections. The Weyl coupling is limited to be $\gamma=\frac{\alpha}{m^2}$ \cite{qc}. An accepted range of 
Weyl's coupling $\gamma $ is the interval $-\frac{1}{16}\gamma \frac{1}{24}$.

%%%%%%%%%%%%%%%%%%%%%%%%%%%%
In probe limit the matter sector is decoupled from the gravity part. So the exact solution of the equations of motion is a planar black hole in $AdS_5$ is given by the following metric:
\begin{eqnarray} ds^{2}=r^{2}(-fdt^{2}+dx^2+dy^2+dz^2)+\frac{dr^{2}}{r^{2}f}.
\end{eqnarray}%
We adopted the coordinates as $x^{\mu}=\{t,r,x,y,z\}$. The single metric function is
\begin{equation*}
f=1-(\frac{h}{r})^{4},
\end{equation*}%
To have temperatures in gravity, we should have a thermalized horizon. It has been provided by the location of horizon at $r=h$. We assume that the temperature of the black hole is same as the CFT temperature.
The Bekenstein-Hawking temperature of the black hole reads as $T=\frac{h}{\pi } $. To support the idea of superconductivity, we mention here that when the horizon becomes cold as much as it falls down to below the specific critical value $T_ {c} $, the system undergoes the phase transition of the second type. The transition is between the normal phase and the superconducting one. At this moment, the scalar field $\psi $ condenses. Such condensation shows that the system has a superfluid phase.\par
%%%%%%%%%%%%%%%%%%%%%%%%%%%%%%%
To quantify our results we start by writing the equations of motion (EOMs) for (\ref{S1}). We are varying the action (\ref{S1}) with respect to the (w.r.t) $A_{\mu},\psi$. 
The resulting Yang-Mills equations
\begin{eqnarray}
D_{\mu}D^{\mu}\psi-m^2\psi=0,  \notag \\
\frac{1}{\sqrt{-g}}\partial_{\mu}(\sqrt{-g}F^{\mu\nu})=i[\psi^{*}D^{\nu}%
\psi-\psi D^{\nu *}\psi^{*}],  \notag
\end{eqnarray}
The first is Klein-Gordon equation for charging scalar field in the presence of electromagnetic fields. The second one is the Maxwell equation for gauge field with the scalar source term. About the mass term $m^2$ as we know, thanks to the Breitenlohner-Freedman bound \cite{Breitenlohner:1982bm}, it should be $m^2\geq m_ {BF} ^2$, where in d-dimensional bulk, $m_{BF}^2=-\frac{(d-1)^2}{4L^2}$. In the next subsection we will study exact analytical solutions of the system near criticality.
%%%%%%%%%%%%%%%%%%%%%%%%
\section{Analytical treatment of field equations}
%%%%%%%%%%%%%%%%%%%%%%%%%
By analytical solution we mean a way to find the critical temperature $T_c$ and condensation expectation value $<\mathcal{O}_{\Delta}>$. There are two major methods to treat the problem analytically:\par
\begin{itemize}
\item Matching method: In which the aim is to match series solutions for fields near the horizon to the AdS asymptotic  solutions monotonically. The method is proposed in \cite{Gregory:2009fj} and widely studied in literature \cite{Bai:2014poa}- \cite{Basu:2011np}. We review it as the following:
consider the set of EOMs for $\{\psi,A_{\mu}=\phi(r)\}$: 
\begin{eqnarray}
&&\psi_{,rr}+p_1(r)\psi_{,r}+q_1(r)\psi=0,\\&&
\phi_{,rr}+p_2(r)\phi_{,r}+q_2(r)\phi=0.
\end{eqnarray}
Where we fix the gauge field $A_{\mu}=\phi(r)dt\in \mathcal{C}$ . But due to the gauge freedom, $\phi^{*}=\phi$. The starting point is to change the radial coordinate to the new dimensions coordinate $z=\frac{h}{r}$. This maps the interval $[h,\infty]\to (0,1] $. So, the set of EOMs reads:
\begin{eqnarray}
&&\psi_{,zz}+p_3(z)\psi_{,z}+q_3(z)\psi=0,\label{eqz1}\\&&
\phi_{,zz}+p_4(z)\phi_{,z}+q_4(z)\phi=0\label{eqz2}.
\end{eqnarray}
Where $\{p_a(z),q_{a}(z)\}_{a=3,4}$ are obtained from $\{p_a(z),q_{a}(z)\}_{a=1,2}$ and some terms of derivatives w.r.t the z. Now we are developing the series solutions for $\psi(z),\phi(z)$ near the horizon as the following:
\begin{eqnarray}
&&\psi(z)=\psi(1)-\psi_{,z}(1)(1-z)+...\label{sol1}\\&&
\phi(z)=\phi(1)-\phi_{,z}(1)(1-z)+...\label{sol2}
\end{eqnarray}
Using an appropriate set of boundary conditions on the fields (which is different in any case) we are able to fix the coefficients in terms of the "physical" parameters $m^2, T, ...$. The next step is we study the far field behavior ($r\to \infty$) or AdS boundary solutions of fields:
\begin{eqnarray}
&&\psi(z)=...+c_{\Delta_{+}}z^{\Delta_{+}}+c_{\Delta_{-}}z^{\Delta_{-}}+...\label{sol3}\\&&
\phi(z)=\phi_{0}+\phi_{1}z+...\label{sol4}
\end{eqnarray}
Here $\Delta_{\pm}$ are conformal dimensions is defined by $\Delta_{\pm}$.
One step towards the final result is to match the set of equations (\ref{sol1},\ref{sol3}) and (\ref{sol2},\ref{sol4}) at an intermediate point $z_m\in(0,1]$. Using this matching we'll find the set of coefficients $\{c_{\Delta_{\pm}},\Delta_{\pm}\}$. It is believed that $c_{\Delta_{\pm}}$ give us the expectation values of the dual operators with a suitable quantization scheme.

\item Variational approach: Since the form of the EOMs are in the self-adjoint (Sturm-Liouville) forms, so always is possible to find the minimum values of eigenvalue and functions \cite{Li:2011xja}-\cite{Lala:2014jca} . We summarize it here :\\ let us to start by the following equivalent form of (\ref{eqz1},(\ref{eqz2}):
\begin{eqnarray}
&&\frac{d}{dz}\Big[P(z)\psi_{,z}\Big]+(\lambda(\mu)+w(z,\phi))\psi=0\label{vareq}
\end{eqnarray}
Near the critical point, we know that $\phi\sim \mu(1-z^b)$, where $\mu$ is the chemical potential and $b$ an exponent which for Weyl corrected HSC is $b=2$. Also the asymptotic form of the solution for $\psi$ is given by $\psi\sim \mathcal{O}_{\pm}>z^{\pm}$. We assume that the complete solution is given by $\psi=F_{\alpha}(z) \mathcal{O}_{\pm}>z^{\Delta_{\pm}}$, where $F_{\alpha}(z)$ is the "trial" function and $\alpha$ denotes a set of variantional parameters which we'll find later. If we substitute these solutions in (\ref{vareq}) we find that, the trial function should satisfy the following Sturm-Liouville equation:
\begin{eqnarray}
&&\frac{d}{dz}\Big[\mathcal{P}(z)F_{\alpha,z}\Big]+\Big(\mathcal{H}(\mu)+\mathcal{W}(z)\Big)F_{\alpha}(z)=0\label{vareq2}
\end{eqnarray}
we know that the minimum of $\mathcal{H}(\mu)$ can be obtained from the extremes of the following functional:
\begin{eqnarray}
Min\{\mathcal{H}(\mu)\}=\frac{\int_{0}^{1}{dz\frac{1}{2}\Big(\mathcal{P}(z)F_{\alpha,z}^2-\mathcal{W}(z)F_{\alpha}^2\Big)}}{\int_{0}^{1}{dz F_{\alpha}^2}}
\end{eqnarray}
The functional gives us the $\mu_{c}$ and the linear relation between $\mu-\mu_{c}$  and $<\mathcal{O}_{\pm}>$.
\end{itemize}

%\textbf{Need corrections}

%%%%%%%%%%%%%%%%%%%%%%%%%
\section{Analytical s-wave holographic superconductors}
%%%%%%%%%%%%%%%%%%%%%%%
To construct a model for s-wave HSC with Weyl corrections let us to start with the action given in (\ref{S1}). As we have been shown that \cite{Momeni:2011ca} in a static gauge $A_{\mu}dr-\varphi(r)dt,\psi=\psi(r)$ the EOMs are written as the following:
\begin{eqnarray}
\psi^{\prime \prime }-\frac{z^4+3}{z(1-z^4)}\psi^{\prime }+(\frac{%
\varphi^2}{h^2(1-z^4)^{2}}-\frac{m^2}{z^2(1-z^4)})\psi=0, \\
(1-24\gamma z^4)\varphi^{\prime \prime }-(\frac{1}{z}+72\gamma
z^3)\varphi^{\prime }-\frac{2\psi^2}{z^2(1-z^4)}\varphi=0,
\end{eqnarray}
Here prime denotes the derivative w.r.t $z$. To fix the boundary conditions appropriately we admit that the scalar field should be remain finite on the horizon and the scalar field must satisfy $\varphi(1)=0,\psi^{\prime }(1)=\frac{ 2}{3}\psi(1)$. To solve the system using the matching method, we write the following solutions in the limit of $z\to0$:
\begin{eqnarray}
&&\varphi\approx \mu-\frac{\rho}{h} z^2, \\&&
\psi \approx\frac{<O_{\Delta_{\pm}}>}{\sqrt{2}h^{\Delta_{\pm}}}%
z^{\Delta_{\pm}}
=\psi^{(1)}z^{\Delta_{+}}+\psi^{(3)}z^{\Delta_{-}},
\end{eqnarray}
Here $\mu$ and $\rho$ define the dual chemical potential and charge density , $\psi^{(1)}$ and $\psi^{(3)}$ are dual to the vacuum expectation values of the boundary operators $O_{\Delta_{\pm}} $ respectively. As we fix the mass term by $m^2=-3$, the conformal dimensions $ \Delta_{\pm}$ are given by
\begin{eqnarray}
\Delta_{\pm}=\{3,1\}.
\end{eqnarray}
%%%%%%%%%%%%%%%%%%%%%%%%%%%%%
At the phase transition when $TT_c$, the system undergoes to criticality and becomes a superconductor. Only at this point the exact solution for gauge field can be written as the following:
\begin{eqnarray}
\varphi=\lambda h_c(1-z^2),
\end{eqnarray}
critical temperature implies critical magnitude of horizon $h_c$ . By plugging this approximate solution in the EOM of a scalar field, we obtain:
\begin{eqnarray}
-\psi^{\prime \prime }+\frac{z^4+3}{z(1-z^4)}\psi^{\prime }-\frac{3}{%
z^2(1-z^4)}\psi=\frac{\lambda^2}{(1+z^2)^2}\psi,
\end{eqnarray}
We define a new parameter $\lambda=\frac{\rho}{h_c^{3}}$. We can solve this equation and find $T_c$. We need to determine the AdS solution of a scalar field, which is already known as the following:
\begin{eqnarray}
\psi(z)= \frac{<O_{\Delta_{\pm}}>}{\sqrt{2}h^{\Delta_{\pm}}}%
z^{\Delta_{\pm}}\Omega(z),
\end{eqnarray}
We normalized the auxiliary function such that it satisfies  $\Omega (0) = 1$. This auxiliary function satisfies the following differential equation:
\begin{eqnarray}
-\Omega^{\prime \prime }+\frac{\Omega^{\prime }}{z}(\frac{z^4+3}{1-z^4}%
-2\Delta_{\pm})+\frac{\Delta_{\pm} ^2 z^4-(\Delta_{\pm}-1)(\Delta_{\pm}-3)%
}{z^2(1-z^4)}\Omega=\frac{\lambda^2}{(1+z^2)^2}\Omega\label{omega-eq},
\end{eqnarray}
when $z\rightarrow 0$, the auxiliary boundary condition is that the quantity $\frac{\Omega^{\prime }}{\xi}$ remains  finite,and consequently  
  $\Omega^{\prime }(0)=0$.

%\subsection{Variational approach}

Note that (\ref{omega-eq}) is in the form of a self-adjoint or Sturm-Liouville eigenvalue problems and it can be written in the following equivalent form adequately:
\begin{equation*}
\frac{d}{dz }[k(z )\frac{d\Omega }{dz }]-q(z)\Omega (z
)+\lambda ^{2}\rho (z)\Omega (z)=0,
\end{equation*}%
with the additional boundary condition:
\begin{equation*}
k(z )\Omega (z )\Omega _{0}^{\prime 1}=0.
\end{equation*}%
This Sturm-Liouville problem can be cast in the form of the variantional problem:
\begin{equation*}
F[\Omega (z )]=\frac{\int_{0}^{1}dz(k(z )\Omega ^{\prime
2}+q(z )\Omega (z )^{2})}{\int_{0}^{1}dz \rho (z )\Omega
(z )^{2}}\label{SL-1}.
\end{equation*}%
As we observe, the parameter $\lambda _{n}$ is given by  $%
\{\lambda _{n}\}_{0}^{\infty }$
 i.e. it is defined by $\lambda _{0}<\lambda
_{n}$. For our model given by (\ref{SL-1}) we obtain:
\begin{eqnarray}
k(z ) &=&z ^{2\Delta _{\pm }-3}(1-z ^{4}), \\
q(z) &=&-z ^{2\Delta _{\pm }-5}\Big(\Delta _{\pm }^{2}z ^{4}-(\Delta
_{\pm }-1)(\Delta _{\pm }-3)\Big), \\
\rho (z) &=&\frac{z ^{2\Delta _{\pm }-3}(1-z^{2})}{1+z
^{2}}.
\end{eqnarray}%

If we fix $\Delta _{+}=3$, then we should satisfy the next boundary condition:
\begin{equation*}
z ^{3}(1-z ^{4})\Omega (z)\Omega _{0}^{\prime 1}=0.
\end{equation*}%
It's clear that on the right boundary point $z=1$ it satisfies but at the left side $z=0$, we must have $$\lim_{z\rightarrow 0}z
^{3}(1-z ^{4})\Omega (z)\Omega ^{\prime }(z)=0$$, it means that the eigenfunction  $%
\Omega (z )$ must be a non-singular function. Consequently $\Omega ^{\prime }(0)=0$. We choose the following polynomial form for the  trial function:

\begin{eqnarray}
\Omega(z)=1-\alpha z^2+\beta z^3\label{trial-1}.
\end{eqnarray}
Now if we substitue (\ref{trial-1}) in (\ref{SL-1}) we obtain:
\begin{eqnarray}
|\lambda_{\alpha,\beta}|^2=\frac{-0.63 \alpha ^2+1.01 \alpha
\beta +2.25 \alpha -0.37 \beta ^2-2 \beta -1.5}{0.015 \alpha
^2+\alpha (-0.02 \beta -0.05)+0.01 \beta ^2+0.03
\beta +0.056},
\end{eqnarray}
This function attains a minimum when $\alpha=-3.92,\beta=4.76$. So we find: 
\begin{eqnarray} |\lambda_{-3.92,4.76}|^2\approx 41.95\label{var-1}, \end{eqnarray}
Using (\ref{var-1}) we are able to compute the \emph{Minimum} of the critical temperature $T_ {c} ^ {Min} $ is given by the following:
\begin{equation*}
T_{c}^{Min}=\frac{h_{c}}{\pi }=\frac{1}{\pi }\sqrt[3]{\frac{\rho }{\lambda }.%
}
\end{equation*}%
Consequently for $\Delta _{+}=3$, $T_{c}^{Min}\approx 0.170845\sqrt[3]{\rho }$, which
is in a great  good agreement with the corresponding numerical value $T_{c}^{Min}=0.170\sqrt[%
3]{\rho }$ for $\gamma =-0.06$ 

Later if we fix $\Delta_{-}=1$, then the following boundary condition must be satisfied:
\begin{eqnarray}
\frac{(1-z^4)}{z}\Omega(z)\Omega^{\prime 1 }_0 = 0.
\end{eqnarray}
The appropriate polynomial form for the trial eigenfunction with the boundary condition $\Omega(0) =0$ is given by:
\begin{eqnarray}
\Omega(z)=az+bz^2+cz^3.
\end{eqnarray}
So, we obtain:

\begin{eqnarray}
\lim_{z\rightarrow0}\frac{1-z^4}{z}(az+bz^2+cz^3)(a+2bz+3c%
z^2)=0.
\end{eqnarray}
Easily we get $a=0$ so we try on $\Omega(z)=bz^2+cz^3$ which it attends to a minimum when $b=0. 465, c=-0.323$. So:

\begin{eqnarray}
|\lambda_{0.465,-0.323}|^2\approx 10.806.
\end{eqnarray}

And consequently for  $\Delta_{-} = 1$, we find $T^{Min}_c\approx 0.21408\sqrt[3]{\rho}$ which it corresponds to the numerical value for $\gamma=0.02$.

%\subsection{Critical exponent $\protect\beta$}

Using the equation of motion of gauge field $\varphi$ at the critical point, we can determine the 
the order parameter. We start by the EOM of $\varphi$ given by the following:
\begin{eqnarray}
(1-24\gamma z^4)\varphi^{\prime \prime }-(\frac{1}{z}+72\gamma
z^3)\varphi^{\prime
}\approx[\frac{<O_{\Delta_{\pm}}>^2}{h^{2\Delta_{\pm}}}
\frac{z^{2\Delta_{\pm}-2}}{1-z^4}\Omega(z)^2]\varphi,
\end{eqnarray}

we define a perturbative parameter $\varepsilon^2=\frac{O_{\Delta_{\pm}}>^2}{ h^{2\Delta_{\pm}}}$ . Since at $T\to T_c$, $\mu\to \mu_c$ , so as a result the condensation is small and we can perform series solution for $\varphi$ as the following:
\begin{eqnarray}
\varphi(z)\sim \mu_c+\varepsilon \chi(z),
\end{eqnarray}
here  $\chi(z)$ is a general function which is restricted to the condition $\chi(0)=1$. It is easy to show that  $\chi(z)$ satisfies the next differential equation appropriately:
\begin{eqnarray}
\chi^{\prime \prime }(z)-\frac{\frac{1}{z}+72\gamma z^3}{1-24\gamma
z^4}\chi^{\prime }(z)=\varepsilon\mu_c\frac{z^{2\Delta_{\pm}-2}}{%
(1-z^4)(1-24\gamma z^4) }\Omega(z)^2.
\end{eqnarray}
where in it:
\begin{eqnarray}
\eta(z)={\frac {\sqrt[4]{-1+24\,\gamma\,{z}^{4}}{\mathrm{e}^{-3\,\sqrt {%
6\gamma }\mathit{arctanh} \left( 2\,\sqrt {6\gamma}{z}^{2} \right) }}}{\xi}%
},  \notag
\end{eqnarray}
We can write the following equation for  $\chi(\xi)$ :
\begin{eqnarray}
\frac{d }{dz} [\eta(z) \frac{d\chi}{dz} ]= -\varepsilon\mu_c\frac{%
z^{2\Delta_{\pm}-3}{\mathrm{e}^{-3\,\sqrt
{6\gamma}\mathit{arctanh} \left( 2\,\sqrt { 6\gamma}{z}^{2}
\right) }}\Omega(z)^2 }{(-1+24\gamma z^4)^{3/4}(1-z^4)}.
\end{eqnarray}
finally we get:
\begin{eqnarray}
\eta(z) \frac{d\chi}{dz}|_{0}^{1} =-\varepsilon\mu_c\int_{0}^{1}\frac{%
z^{3}{\mathrm{e}^{-3\,\sqrt {6\gamma}\mathit{arctanh} \left(
2\,\sqrt { 6\gamma}{z}^{2} \right) }}(1-\alpha z^2)^2
}{(-1+24\gamma z^4)^{3/4}(1-z^4)}dz,
\end{eqnarray}
Using the trial function $\Omega(z)=1-\alpha z^2$, and if we fix $ \Delta_{+} =3$, we find that in the vicinity of the $z=0$, we have :
\begin{eqnarray}
\varphi\approx \mu-\frac{\rho}{h} z^2\approx\mu_c+\varepsilon(\chi(0)+%
\chi^{\prime }(0)z+\frac{1}{2}\chi^{\prime \prime 2}+...).
\end{eqnarray}
by finding the corresponding coefficients of diffrent powers of  $z^0$ we get:
\begin{eqnarray}
\mu-\mu_c\approx\varepsilon\chi(0),
\end{eqnarray}
and for  $z^1$we obtain  $\chi^{\prime }(0)=0$ and we solve to find $\chi(z)$ :
\begin{equation}
\chi(z)=-\varepsilon \mu_c\{c^{\prime }+\frac{cz^2}{2}+9c\gamma z^4+\{%
\frac{-1}{24}+c\gamma(1+108\gamma)\}z^6+O(z^7)\}.
\end{equation}
Here, $\{c,c^{\prime }\}$ is a pair of the integration constants. Consequently we  find$%
\chi(0)\approx-\varepsilon \mu_c c^{\prime }$. Further we have
\begin{eqnarray}
\mu-\mu_c\approx- \mu_c c^{\prime 2}, \\
<O_{\Delta_{\pm}}>\approx \frac{h^3}{\sqrt{-c^{\prime }\mu_c}}\sqrt{\mu-\mu_c%
}.
\end{eqnarray}
We find the famous critical exponent $\frac{1}{2}$ for the condensation value and furthermore we know that $ \mu-\mu_c$ qualitatively match the numerical curves . Furthermore we have been proven a linear relation always exists between the dual charge density $\rho$ and the dual chemical potential ( $\mu-\mu_c$) .
%%%%%%%%%%%%%%%%%%%%%%%%%%%%%%%%
\section{St\'{u}ckelberg  's correction}
%%%%%%%%%%%%%%%%%%%%%%%%%%%%%%%%%
This section is devoted to study the effects of non linearity due to the St\'{u}ckelberg   term in Weyl's corrected s-wave HSC. The modified action which should be replaced in place of (\ref{S1}) is given by the following form \cite{Ma:2011zze}:
\begin{eqnarray}
&&S=\int dtd^{4}x\sqrt{-g}\Big(R+\frac{12}{l^2}-\Big(\frac{1}{4}(F^{\mu\nu}F_{\mu\nu}-4\gamma
C^{\mu\nu\rho\sigma}F_{\mu\nu}F_{\rho\sigma})\\&&\nonumber+ \frac{\nabla_{\mu}\psi%
\nabla^{\mu}\psi}{2}+\frac{m^2\psi^2}{2}+\frac{1}{2}F(\psi)A_{\mu}A^{\mu}\Big)\Big),  \label{S2}
\end{eqnarray}
Here we add a new mass term to the gauge field, the Stuckelberg potential function is given by:
\begin{eqnarray}  \label{F}
F(\psi)=\psi^2+c_{\alpha}\psi^{\alpha}+c_4\psi^4.
\end{eqnarray}
where $3\leq\alpha\leq4$ and $c_4,c_{\alpha}$ are constants of order unit.
The action has been written in the "God-given" units 
 $2\kappa^2=1,$ and we put AdS radius $L=1$,charge $e=1$, $F_{\mu\nu}=\partial_{\mu}A_{\nu}-%
\partial_{\nu}A_{\mu}$. 
 In the
probe limit,in the regime of decoupling of gravity from the matter fields, 
 the exact solution for Einstein-Maxwell system is given by the following planar 
 AdS-Schwarzschild black hole :
\begin{eqnarray}
ds^2=r^2(-fdt^2+dx^idx_i)+\frac{dr^2}{r^2f} ,  \label{g}
\end{eqnarray}
where
\begin{eqnarray}
f=1-(\frac{h}{r})^4,
\end{eqnarray}
The black hole is in thermal equilibrium with the CFT boundary, Our aim is to study the analytical properties of this system near criticality \cite{Momeni:2012uc}. We choose the fields as given by $ \psi=\psi(r),A_{t}=\varphi(r)$ and we rewrite the EOMs in the dimensionless forms as the following:
\begin{eqnarray}
\psi^{\prime \prime }+\Big(\frac{f^{\prime }}{f}+\frac{5}{r}\Big)%
\psi^{\prime }+\frac{\phi^2}{2r^4 f^2}\frac{dF}{d\psi}-\frac{m^2\psi}{r^2 f}%
=0,  \label{eom1} \\
\ \ \Big(1-\frac{24\gamma h^4}{r^4}\Big)\phi^{\prime \prime }+\Big(\frac{3}{r%
}+\frac{24\gamma h^4}{r^5}\Big)\phi^{\prime }-\frac{F}{r^2 f}\phi=0,
\label{eom2}
\end{eqnarray}
where \texttt{prime}$\equiv \partial_r$. In terms of the $\xi$,
Eqs. (\ref{eom1}), (\ref{eom2}) become
\begin{eqnarray}
\frac{d}{d\xi}\Big(\xi^2\frac{d\psi}{d\xi}\Big)-\frac{\xi(5-\xi^4)}{1-\xi^4}%
\frac{d\psi}{d\xi}+\frac{\phi^2\xi^2}{2h^2(1-\xi^4)^2}\frac{dF(\psi)}{d\psi}-%
\frac{m^2}{1-\xi^4}\psi=0,  \label{eom11} \\
\ \ (1-24\gamma\xi^4)\frac{d}{d\xi}\Big(\xi^2\frac{d\phi}{d\xi}\Big)%
-3\xi(1+8\gamma\xi^4)\frac{d\phi}{d\xi}-\frac{F(\psi)\phi}{1-\xi^4}=0.
\label{eom22}
\end{eqnarray}
The appropriate set of boundary conditions in bulk is given by  $\varphi(1)=0,\psi^{\prime }(1)=\frac{2}{3}\psi(1)$. So the AdS solutions for the fields which are valid till $r\to\infty$ are:
\begin{eqnarray}
\varphi\approx \mu-\frac{\rho}{h} z^2, \\
\psi \approx\epsilon z^{\Delta_{\pm}}
=\psi^{(1)}z^{\Delta_{+}}+\psi^{(3)}z^{\Delta_{-}},
\label{aproxpsi}
\end{eqnarray}
where we define the smallness parameter \cite{Kanno:2011cs},
\begin{eqnarray}
\epsilon=\frac{<O_{\Delta_{\pm}}>}{\sqrt{2}h^{\Delta_{\pm}}},
\label{epsilon}
\end{eqnarray}
As usual terminology, $\mu,\rho$ are dual  densities  for  CFT, $\psi^{(1)},\psi^{(3)}$ are dual to the 
expectation values of relevant quantum operator $\mathcal{O}_{1,3}$ and we adopted the mass term such that :
\begin{eqnarray}
\Delta_{\pm}\equiv \Delta=\{3,1\}.
\end{eqnarray}

%%%%%%%%%%%%%%%%%%%%%%%%%%%%%%%%%%%%%%%%%%%%%%%%%%%%%%%%%%%%%%%%%%%%%%%%%%%%%%%%

%\subsection{Analytical results for the condensation and critical temperature}

%%%%%%%%%%%%%%%%%%%%%%%%%%%%%%%%%%%%%%%%%%%%%%%%%%%%%%%%%%%%%%%%%%%%%%%%%%%%%%%%
As we know  the solution of Eq.(\ref{eom22}) in the vicinity of the criticality  $T_c$ is given by:
\begin{eqnarray}
\phi=\lambda h_c(1-z^2),
\end{eqnarray}
Consequently, when  $T\rightarrow T_c$,
the EOM given by (\ref{eom11}) changes to  the following diffrential equation:
\begin{eqnarray}
-\psi^{\prime \prime }+g(z)\psi^{\prime }+\frac{m^2}{z^2(1-z^4)}\psi=%
\frac{\lambda^2} {2(1+z^2)^2}\frac{dF}{d\psi},  \label{psi}
\end{eqnarray}

where $\lambda=\frac{\rho}{h_c^{3}}$, and we defined:
\begin{equation*}
g(z)=\frac{z^4+3}{z(1-z^4)}.
\end{equation*}
By using the Poincare's asymptotic method, we are writing the following exact solution for the field:
\begin{eqnarray}
\psi(z)= \epsilon z^{\Delta}\Omega(z),\label{psi2}
\end{eqnarray}
where the normalized auxiliary function  $\Omega$ satisfies $\Omega (0)
= 1$. We substitue (\ref{psi2}) in (\ref{psi}) and we obtain:
\begin{eqnarray}
-\Omega^{\prime \prime }+(g(z)-\frac{2\Delta}{z})\Omega^{\prime }+(\frac{%
\Delta g(z)}{z}-\frac{\Delta(\Delta-1)}{z^2}+\frac{m^2}{z^2(1-z^4)}%
)\Omega=\frac{\lambda^2
z^{-\Delta}}{2(1+z^2)^2}\frac{dF}{d\psi}|_{ \psi(z)=
\epsilon z^{\Delta}\Omega(z)},  \label{omega}
\end{eqnarray}
To have regular solution when  $\xi\rightarrow 0$, we must control the term $\frac{\Omega^{\prime }}{\xi}$ to remains  finite, in particular case we suppose that $%
\Omega^{\prime }(0)=0$.

%%%%%%%%%%%%%%%%%%%%%%%%%%%%%%%%%%%%%

%\subsection{Variational approach}

%%%%%%%%%%%%%%%%%%%%%%%%%%%%%%%%%%%%%%%%

The Sturm-Liouville formulation of the problem is given by the following second order , self-adjoint differential equation :
\begin{eqnarray}
\frac{d }{dz}[k(\xi) \frac{d\Omega}{dz} ] - q(z)\Omega(z) + \frac{%
\lambda^2\rho(z)}{2}\Phi^{\prime }(\Omega(z)) = 0,
\end{eqnarray}
which is constrained to the following auxiliary boundary condition:
\begin{eqnarray}
k(z)\Omega(z)\Omega^{\prime 1 }_0 = 0.  \label{bc}
\end{eqnarray}
The variational problem  can be cast to the functional minimization problem:
\begin{eqnarray}
F [\Omega(z)] = \frac{\int ^1 _0 dz(k(z)\Omega ^{\prime 2 }+
q(z)\Omega(z) ^2)}{\int ^1 _0 dz\rho(z)\Phi(\Omega(z)) }.
\label{functional}
\end{eqnarray}
For Eq.(\ref{omega}) we  have:
\begin{eqnarray}
&&k(z)=z^{2\Delta-3}(1-z^4), \\
&&q(z)=-k(z)(\frac{\Delta g(z)}{z}-\frac{\Delta(\Delta-1)}{z^2}+%
\frac{m^2}{z^2(1-z^4)}), \\
&&\rho(z)=z^{\Delta-3}\frac{1-z^2}{1+z^2}, \\
&&\Phi(\Omega(z))\equiv F(\Omega(z))= ( \epsilon z^{\Delta}\Omega(z))^2%
\Big(1+c_4 ( \epsilon z^{\Delta}\Omega(z))^2+c_{\alpha} (
\epsilon z^{\Delta}\Omega(z))^{\alpha-2}\Big).
\end{eqnarray}
%Where F is defined by (\ref{F}).
%We will follow this method in next section.

%%%%%%%%%%%%%%%%%%%%%%%%%%%%%%%%%%%%%%%%%%%%%%%%

%\subsection{Analytical results for $\Delta=3$}

%%%%%%%%%%%%%%%%%%%%%%%%%%%%%%%%%%%%%%%%%%%%%%%%%%%
For conformal dimension given by  $\Delta=3$, the boundary condition (\ref{bc}) reads as the following:
\begin{eqnarray}
z^3(1-z^4)\Omega(z)\Omega^{\prime 1 }_0 = 0.
\end{eqnarray}
By using an appropriate trial function  
%\begin{figure}
%\centering
%\includegraphics[width=10cm,angle=0] {1}% scale goes from 0 to 1.
%\caption{Variation of the functions $k(\xi),q(\xi),\rho(\xi)$ for the case $\Delta_{+}=3$}
%\label{1.eps}
%\end{figure}

\begin{eqnarray}
\Omega(z)=1-\beta z^2.
\end{eqnarray}
We obtain:
\begin{eqnarray}
\lambda^2(\beta,m^2,\alpha)=-\frac{20.4855 \left( m^2 \beta ^2-3 m^2
\beta +3m^2+6.8 \beta ^2-22.5 \beta +18\right)}{\left( \beta
^2-2.96617 \beta +2.41891\right) \epsilon ^2}.  \label{integral}
\end{eqnarray}
The minimum of $%
\lambda^2(\beta,m^2,\alpha)$ located  at $\beta=0.304936$, with the minima magnitude as given by:

\begin{eqnarray}
|\lambda_{min}|^2=\frac{12.7445 \left(2.17818 m^2+11.7713\right)}{\epsilon ^2%
}.
\end{eqnarray}

Consequently the minimum  of the critical temperature $T_{c}^{Min}$ is obtained:
\begin{eqnarray}
&&T_{c}^{Min}=\frac{h_{c}}{\pi }=\frac{1}{\pi }\sqrt[3]{\frac{\rho \epsilon }{%
\sqrt{12.7445\left( 2.17818m^{2}+11.7713\right) }}}.
\end{eqnarray}%
One significant result is , we observe that the mass of the scalar field has a lower bound:
\begin{equation*}
m^{2}>m_{c}^{2},
\end{equation*}%
where $m_{c}^{2}=-5.40417$ which this bound lives in the range of the BF bound. \par

Similarly for $m^2=-3$ we find:
\begin{eqnarray}
T^{Min}_c=0.158047\sqrt[3]{\rho\epsilon }.
\end{eqnarray}
which is comparable to the numerical value of the critical temperature  for Weyl's coupling
$\gamma=-0.06$ is given by the following \cite{Ma:2011zze}:
\begin{equation*}
T_c^{Min-Numerical}\approx 0.170\sqrt[3]{\rho}.
\end{equation*}
Our analytical result is obtained as follows:
\begin{equation*}
T_c^{Min-Analytical}\approx0.158047\sqrt[3]{\rho\epsilon }.
\end{equation*}
can be read-off at the lower bound for the numerical estimation. if  we put $\epsilon\approx1.25$, the results coincide to  the numerical estimation \cite{Ma:2011zze}

%%%%%%%%%%%%%%%%%%%%%%%%%
%\subsection{Linear relation between $<O_{\Delta }>$ and the chemical
%potential}

Is it possible also to present the linear relation between $<O_{\Delta_{\pm} }>$
and the chemical potential. using the critical solution of the following diffrential equation :
\begin{equation*}
\phi ^{\prime \prime }+s(z )\phi ^{\prime }=\frac{(\epsilon z ^{\Delta
}\Omega (z ))^{2}\Big(1+c_{4}(\epsilon z^{\Delta }\Omega (z
))^{2}+c_{\alpha }(\epsilon z ^{\Delta }\Omega (z ))^{\alpha -2}\Big)}{%
z ^{2}(1-z ^{4})(1-24\gamma z^{4})}\phi.
\end{equation*}%
Here
\begin{equation*}
s(z )=-\frac{(1+72\gamma z ^{4})}{z (1-24\gamma z ^{4})}.
\end{equation*}%
We must keep only those terms of series which have the order of magnitude as $\epsilon ^{4}$(
$3<\alpha <4$). We put $c_{4}\cong 0$, by  keeping the term $
c_{\alpha }$ , we write the following approximate solution near the criticality $T_c$:
\begin{equation*}
\phi (z)\approx \mu _{c}+\epsilon \chi (z),
\end{equation*}%
where $\chi (z )$ is the trial function restricted to the  auxiliary condition $%
\chi (0)=1$. The differential equation for $\chi $ is given by the following:
\begin{eqnarray}
&&\chi ^{\prime \prime }(z)+s(z)\chi ^{\prime }(z )\approx \mu _{c}%
\frac{\epsilon (z ^{\Delta }\Omega (z))^{2}\Big(1+c_{\alpha }(\epsilon
z ^{\Delta }\Omega (z))^{\alpha -2}\Big)}{z ^{2}(1-z
^{4})(1-24\gamma z ^{4})}  \label{chi}
\end{eqnarray}

The exact solution for $\chi (z)$ reads
\begin{eqnarray}
&&\chi ^{\prime -\int s(z )dz }(C+\epsilon \int j(z )e^{\int
s(z )dz}dz ).
\end{eqnarray}%
Where
\begin{eqnarray}
&&j(z)=\mu _{c}\frac{(z ^{\Delta }\Omega (z
))^{2}\Big(1+c_{\alpha }(\epsilon z ^{\Delta }\Omega (z
))^{\alpha -2}\Big)}{z ^{2}(1-z ^{4})(1-24\gamma z
^{4})}\newline ,\ \ e^{\int s(z )dz}=\frac{-1+24\gamma z
^{4}}{z}  \label{chi1}.
\end{eqnarray}%
Finally  we obtain:
\begin{equation*}
\chi ^{\prime }(z)=\frac{z}{24\gamma z ^{4}-1}(C+\epsilon \int \frac{%
j(z)(24\gamma z ^{4}-1)}{z}dz ).
\end{equation*}%
and consequently we find:
\begin{eqnarray}
\chi (0)=\frac{\sqrt{6}\pi C}{48\sqrt{\gamma }}\label{chi0}.
\end{eqnarray}

If we expand  in series the $\phi$ in the vicinity of $z=0$ we find:
\begin{eqnarray}
\phi\approx \mu-\frac{\rho}{h}z^2\approx
\mu_c+\epsilon(\chi(0)+\chi^{\prime }(0)z+\frac{1}{2}\chi^{\prime
\prime 2}(0)z^2+...).
\end{eqnarray}
By comparing the coefficents of term with  $z^0$ we obtain:
\begin{eqnarray}
\mu-\mu_c\approx \epsilon\chi(0).
\end{eqnarray}
and using (\ref{chi0}) we have:
\begin{eqnarray}
\mu-\mu_c\approx \epsilon\Big( \frac{ \sqrt{6} \pi C }{48 \sqrt{\gamma } }%
\Big).
\end{eqnarray}
consequently:
\begin{eqnarray}
\mu-\mu_c\approx <O_{\Delta}>\Big( \frac{ \sqrt{6} \pi C }{48
\sqrt{\gamma } }\Big).
\end{eqnarray}
It implies the linear relation between $<O_{\Delta}>$ and the chemical potential .

%%%%%%%%%%%%%%%%%%%%%%%%
\section{Buliding a p-wave holographic superconductor}
%%%%%%%%%%%%%%%%%%%%%%%%

%\subsection{ Weyl corrected P-wave superconductors}
This section is devoted to constructing a model for the p - wave HSC with Weyl corrections. What we need is to replace the Maxwell field with $
SU (2) $ Yang-Mills (YM) field $A_{\mu } ^ {a} $. To include the effects of Weyl corrections, we will work in a five dimensional space-time. So, the holographic picture is observed on $1+3$ dimensions. We propose the following model for p-wave \cite{Momeni:2012ab}, \cite{Momeni:2013fma}
: 
\begin{equation}
S=\int dtd^{4}x\sqrt{-g}\{\frac{1}{16\pi
G_{5}}(R+12)-\frac{1}{4g^{2}}F_{\mu \nu }^{a}F^{a\mu \nu }+\gamma
C^{\mu \nu \rho \sigma }F_{\mu \nu }^{a}F_{\rho \sigma }^{a}\}.
\end{equation}
where $G_{5}$ denotes the effective five dimensional Newtonian constant, $g$ is the
Yang-Mills coupling constant. To satisfy AdS/CFT, we need to specify the negative cosmological constant is given by  $\frac{12}{L^2},\ \ l=1$.  With the same terminology as s-wave, we define the  strength tensor for  
the non-Abelian gauge fields $A_{\mu}^{a}$ in the following form:

\begin{equation}
F_{\mu \nu }^{a}=\partial _{\mu }A_{\nu }^{a}-\partial _{\nu }A_{\mu
}^{a}+\varepsilon ^{abc}A_{\mu }^{b}A_{\nu }^{c},\{a=1,2,3,\mu
=0,1,2,3\}.
\end{equation}%
The Weyl's term is coupled to the strength tensor through the coupling constant $\gamma$ in the same manner as a s - wave. The following planar version of AdS-Schwarzschild black hole solves EOMs in probe limit:
\begin{equation}
ds^{2}=r^{2}(-fdt^{2}+dx^2+dy^2+dz^2)+\frac{dr^{2}}{r^{2}f}\label{metric3}.
\end{equation}%
where we have:
\begin{equation}
f=1-(\frac{r_{+}}{r})^{4}
\end{equation}%
We assume that the horizon is located at $r=r_{+}$. Also to have thermal behavior, we define the Smarr-Bekenstein-Hawking
temperature of the black hole by $%
T=\frac{r_{+}}{\pi }$. EOMs are given by 
Euler-Lagrange equations or using the  generalized Yang-Mills equation :
\begin{equation}
\nabla _{\mu }\left( F^{a\mu \nu }-4\gamma C^{\mu \nu \rho \sigma
}F_{\rho \sigma }^{a}\right) =-\epsilon _{bc}^{a}A_{\mu }^{b}F^{c\mu
\nu }+4\gamma C^{\mu \nu \rho \sigma }\epsilon _{bc}^{a}A_{\mu
}^{b}F_{\rho \sigma }^{c}. \label{GYME1}
\end{equation}%
For metric given by (\ref{metric3}), the non-zero components of the Weyl tensor  $C_{\mu \nu \rho \sigma }$ are as the following:
\begin{equation}
C_{0i0j}=f(r)r_{+}^{4}\delta _{ij},~~C_{0r0r}=-\frac{3r_{+}^{4}}{r^{4}}%
,~~C_{irjr}=-\frac{r_{+}^{4}}{r^{4}f(r)}\delta
_{ij},~~C_{ijkl}=r_{+}^{4}\delta _{ik}\delta _{jl}.  \label{WeylT}
\end{equation}
To make the problem simpler we assume that the Yang-Mills gauge field has the following form:
\begin{equation*}
A=\varphi (r)\sigma ^{3}dt+\psi (r)\sigma ^{1}dx,
\end{equation*}%
here $\sigma ^ {I} $ denotes Pauli's matrices. The condensation scenario of $\psi (r) $ breaks the $%
SU (2) $ symmetry it implies that the system undergoes to the second order phase transition with superconducting ultimate phase. From the gauge/gravity dictionary, we label the gauge function $\psi (r) $as the dual of the CFT vector operator $J_ {x} ^ {1} $ where we choose $x$ axis as the preferred direction.\par
 The resulting
Yang-Mills equations for (\ref{metric3}) are given by:
\begin{eqnarray}
&&\left( 1-\frac{24\gamma r_{+}^{4}}{r^{4}}\right) \varphi ^{\prime \prime
}+\left( \frac{3}{r}+\frac{24\gamma r_{+}^{4}}{r^{5}}\right) \varphi
^{\prime }-\left( 1+\frac{8\gamma r_{+}^{4}}{r^{4}}\right) \frac{\psi
^{2}\varphi }{r^{4}f} =0,  \label{EOMr1} \\
&&\left( 1-\frac{8\gamma r_{+}^{4}}{r^{4}}\right) \psi ^{\prime \prime }+\left[
\frac{3}{r}+\frac{f^{\prime }}{f}-\frac{8\gamma r_{+}^{4}}{r^{4}}\left( -%
\frac{1}{r}+\frac{f^{\prime }}{f}\right) \right] \psi ^{\prime }+\left( 1+%
\frac{8\gamma r_{+}^{4}}{r^{4}}\right) \frac{\varphi ^{2}\psi
}{r^{4}f^{2}} =0,  \label{EOMr2}
\end{eqnarray}%
where the prime in all functions, denotes derivative with respect to $r$. By going to the dimensionless coordinate $z$ , we change the  (\ref{EOMr1}) and (\ref{EOMr2}) to the following system to be solved numerically and analytically:
\begin{eqnarray}
&&\left( 1-24\gamma z^{4}\right) \varphi ^{\prime \prime }-\frac{1}{z}\left(
1+72\gamma z^{4}\right) \varphi ^{\prime }-\left( 1+8\gamma z^{4}\right)
\frac{\psi ^{2}\varphi }{f} =0,  \label{EOMz1} \\
&&\left( 1-8\gamma z^{4}\right) \psi ^{\prime \prime }+\left[ -\frac{1}{z}+%
\frac{f^{\prime }}{f}-8\gamma z^{4}\left( \frac{3}{z}+\frac{f^{\prime }}{f}%
\right) \right] \psi ^{\prime }+\left( 1+8\gamma z^{4}\right)
\frac{\varphi ^{2}\psi }{f^{2}} =0 , \label{EOMz2}
\end{eqnarray}%
The asymptotic AdS boundary
conditions , i.e. $z\rightarrow 0$, are given by :
\begin{eqnarray}
\varphi &\simeq &\mu -\rho z^{2} \\
\psi &\simeq &\psi ^{(0)}+\psi ^{(2)}z^{2},
\end{eqnarray}%
where for the renormaliztion reason we put $\psi ^ {(0)} =0, \ \ \mathcal{O}_{\Delta_{1}}=0$. The numerical results have been shown in the next table, which it implies the superconductivity in this toy model for different values of Weyl's coupling:

\begin{table}[th]
\begin{center}
\begin{tabular}{|c|c|c|c|c|c|c|}
\hline
$~~\gamma:\ \ \texttt{Weyl coupling}~~$ & ~~$-0.06$~~ & ~~$-0.04$~~ & ~~$-0.02$~~ & ~~$0$~~ & ~~$0.02$~~
& ~~$0.04$~~ \\ \hline
~~$\frac{T_{c}}{\rho^{1/3}}:\ \ \texttt{Normalized critical temperature}$~~ & ~~$0.1701$~~ & ~~$0.1774$~~ & ~~$%
0.1869$~~ & ~~$0.2005$~~ & ~~$0.2239$~~ & ~~$%
0.3185$~~ \\ \hline
\end{tabular}%
\end{center}
\caption{The critical temperature $T_{c}$ which has been obtained by numerical methods. We adopted the fields as $\varphi(1)=\varphi^{\prime }(0)=\psi^{\prime }(0)=\psi^{\prime }(1)=0$}
\label{Tc}
\end{table}

%\subsection{Numerical treatment}

Furthermore we demonstrated that the condensation happened as a function of temperature for the dual operator$ J^1 _x >$. We observed that the condensation of $ <J^1 _x >|_{T\to 0}\sim J_0$. Also $T_c$ is increased when the $\gamma$ varies in the range of $-0.06$ to $0.04$ and particularly when $\gamma 0 $ , $T_c$ decreases and the condensation becomes harder.

%\subsection{Analytical treatment: variational}

%\subsection{Critical temperature $T_C^{Min}$}

Analytical method is also provided. If we suppose that the fields have the following critical solutions:
\begin{equation*}
\psi (z)=0,\varphi (z)=\lambda r_{+c}(1-z^{2})
\end{equation*}%
where $\lambda =\frac{\rho }{r_{+c}^{3}}$, $r_{+c}$ denotes the radius of the
horizon at c riticality $T=T_{c}$ we have the following EOM for $\psi $ :
\begin{equation}
z^{2}\frac{d}{dz}((\frac{(1-z^{4})}{2g^{2}z}+4\gamma r_{+c}^{3}z^{3})\frac{%
d\psi }{dz})+\lambda ^{2}[r_{+c}^{3}(\frac{z}{2g^{2}}+4\gamma z^{5})\frac{%
1-z^{4}}{1+z^{4}}]\psi =0
\end{equation}%
We suggest the following Poincare's solution:
\begin{equation}
\psi (z)=\frac{\langle J_{x}^{1}\rangle }{r_{+}}z^{2}F(z)
\end{equation}%
with the auxiliary  boundary conditions  $F(0)=1,F^{\prime }(0)=0$. So, the auxiliary function $f(z)$ satisfies the following differential equation:

\begin{equation}
\frac{d}{dz}(k(z)\frac{dF(z)}{dz})-p(z)F(z)+\lambda ^{2}q(z)F(z)=0
\end{equation}

where:

\begin{eqnarray}
k(z)=\frac{z^3(1-z^4)}{2g^2 }+4\gamma r_{+c}^3 z^7 \\
p(z)=-2z^5(-\frac{2}{g^2}+16\gamma r_{+c}^3) \\
q(z)=r_{+c}^3z^2(\frac{r_{+c}z}{2g^2}+4\gamma z^5)\frac{1-z^4}{1+z^4}
\end{eqnarray}

The minimum of the  eigenvalue $\lambda $ is obtained using the following functional:
\begin{equation}
\lambda ^{2}=\frac{\int_{0}^{1}(k(z)F^{\prime 2}+p(z)F(z)^{2})dz}{%
\int_{0}^{1}q(z)F(z)^{2}dz}
\end{equation}%
Using a trial function $F(z)=1-\alpha z^{2}$ we find:
\begin{equation}
\lambda _{\alpha }^{2}=\frac{2g^{2}\left( -1.6r_{+c}^{3}\alpha ^{2}\gamma
+8r_{+c}^{3}\alpha \gamma -5.33h_{c}^{3}\gamma +\frac{0.53\alpha ^{2}%
}{g^{2}}-\frac{\alpha }{g^{2}}+\frac{0.67}{g^{2}}\right) }{%
r_{+c}^{3}\left( \alpha ^{2}\left( -0.22\gamma g^{2}-0.015\right)
+\alpha \left( 0.63\gamma g^{2}+0.053\right) -0.48\gamma
g^{2}-0.057\right) }
\end{equation}%
Which attains a local minimum when  $\alpha =0.30$,  the minimum value of the
critical temperature is given by:
\begin{equation}
T_{c}^{Min(\pm )}=0.27\sqrt[3]{\frac{-0.13\pm 0.31g^{2}\sqrt{\frac{%
1.90\gamma g^{2}\rho ^{2}\left( 1.08\gamma g^{2}+0.13\right)
+0.17}{g^{4}}}}{\gamma g^{2}}}
\end{equation}

If we pass to the regime of the strong coupling  of the YM theory, it is possible to perform a series expansion for  $\frac{1}{g}$ as:
\begin{equation*}
T_{c}^{Min(+)}\approx 0.194\rho ^{\frac{1}{3}}+\frac{%
0.15(-0.13\gamma +0.03\gamma \rho )}{\gamma ^{2}\rho ^{\frac{%
2}{3}}g^{2}}
\end{equation*}%
In probe for such large value of the YM \cite{Gorski:1983pp} coupling , the analytic value of the leading order $T_{c}^{Min}\approx
0.194\rho ^{\frac{1}{3}}$ obeys the rule $T_{c}\propto \rho ^{\frac{1}{3}}$ and it is considered as the lower bound of $%
T_ {c} $ according to the numerical results. 
The minimum numeric value of $T_ {c} $ read :
\begin{equation*}
T_{c(numeric)}^{Min(+)}\approx 0.1701\rho ^{\frac{1}{3}}
\end{equation*}%
As we see,  the analytic is in good agreement with  the numerical
estimation.

%\textbf{completed}
%\subsection{Relation of $<J_x^1> -(\protect\mu-\protect\mu_c)$}
Similarly, we can investigate the criticality for the scalar field. This latest result gives us the relation between the condenser and the difference between the normal and critical values of the chemical potential $\mu-\mu_c$. Let us start by the EOM of gauge field $\varphi$ as the following:

\begin{eqnarray}
\frac{d}{dz}((-\frac{1}{2z g^2}+12\gamma z^3)\frac{d\varphi}{dz})+(\frac{z}{%
r_c})^2(\frac{ z}{2(1-z^4)g^2 }+\frac{4\gamma z^5}{1-z^4 })(\frac{<J^1 _x>}{%
r_{+}} )^2F(z)^2\varphi=0
\end{eqnarray}
With the same logic as we use it before, we expand the gauge field in criticality in the following form:
\begin{eqnarray}
\varphi=\mu_c+<J^1 _x>\chi(z)+ ...
\end{eqnarray}
The appropriate boundary condition is written as $\chi(1)= 0$ . If we substitute this approximated solution, we find the following differential equation for the auxiliary function $\chi(z)$:
\begin{eqnarray}
\frac{d}{dz}((-\frac{1}{2z g^2}+12\gamma z^3)\frac{d\chi(z)}{dz})+(\frac{z}{%
r_{+c}})^2(\frac{ z}{2(1-z^4)g^2 }+\frac{4\mu_c\gamma z^5}{1-z^4 })\frac{<J^1 _x>}{%
r_{+c}^2}F(z)^2=0\label{chi2}
\end{eqnarray}
It is completely straightforward to solve (\ref{chi2}) using integration, we obtain:
\begin{eqnarray}
(-\frac{1}{2z\zeta^2}+12\gamma z^3)\frac{d\chi(z)}{dz}=-\mu_c\frac{<J^1 _x>}{%
r_{+c}^4}\int z^2(\frac{ z}{2(1-z^4)g^2 }+\frac{4\gamma z^5}{1-z^4 })F(z)^2 dz
\end{eqnarray}
Since the $\chi(z)$ is considered as a regular function in the vicinity of the AdS boundary point, we suppose that $\chi^{\prime } (0) =0$ . The adequate form for the trial function is given by $%
F(z)=(1-z^2)(1-\alpha z^2) $. We substitute it in the integral, and meanwhile we expand $\varphi(z)$ in the following series form:
\begin{eqnarray}
\varphi(z)\sim\mu-\rho z^2\approx\mu_c+<J^1 _x>(\chi(0)+\chi^{\prime
}(0)z+...)
\end{eqnarray}
The rest is so easy to compare the coefficients of $z^0$ of both sides of this equation. We obtain:

\begin{eqnarray}
&&\mu-\mu_c\approx\frac{(<J^1 _x>)^2 \mu_c}{34560 r_{+}^4 \gamma ^2 g ^4}\Big( 3003.22
\gamma ^{3/2} g ^3 \left(8 \gamma g ^2+1\right) \text{Li}_{2}\Big[\frac{12
\sqrt{\gamma } g }{12\sqrt{\gamma } g -2.44949}\Big]\\&&\nonumber+(9047.79+6359.31 )
\gamma ^2 g ^4\Big)
\end{eqnarray}
When $\alpha=0.305$  we have a minima ($%
\text{Li}_{2}(z)$denotes the  polylogarithm function).\newline
We obtained the critical exponent  $\frac{1}{2}$which is in great agreement with the mean field theory (GL) and also the numerical results.

%%%%%%%%%%%%%%%%%%%%
%p-wave-IJMPA
%%%%%%%%%%%%%%%%%%%%
%%%%%%%%%%%%%%%%%%%%%%%%%%%%%%%%%%%%%%%%%%%%%%%%%%%%%%%%%%%%%%%%%%%%%%%%%%%

%\subsection{Analytical p-wave superconductors with Weyl corrections: matching method}

%%%%%%%%%%%%%%%%%%%%%%%%%%%%%%%%%%%%%%%%%%%%%%%%%%%%%%%%%%%%%%%%%%%%%%%%%%%%%%

%%%%%%%%%%%%%%%%%%%%%%%%%%%%%%%%%%%%%%%%%%%%%%%%%

%\subsection{Existence of a Superconductor phase}

%%%%%%%%%%%%%%%%%%%%%%%%%%%%%%%%%%%%%%%%%%%%%%%%%%
It is possible to study p-wave superconductors by matching method. 
But before, it is an illustrative example to show that whether the p-wave model is stable or not?. Let us start by the EOMs given in (\ref{EOMr1}) and (\ref{EOMr2}). In the normal phase, when the system did not enter to the superconductivity, $\psi=0$ and we have the following exact solution (trivial) for EOMs:
\begin{eqnarray}
\phi = \phi_0(r) &=& \frac{\rho}{r_+^2} \left ( 1 - \frac{r_+^2}{r^2} \right
),\ \ \rho=\mu r_{+}^2,
\end{eqnarray}
It is verified by direct substitution of the solutions in the field equations. The set of solutions describes the normal phase, no condensation phase of the system. Now, we must prove that due to the instability, there exists a unique physical state of the system which when the system reaches it, the system becomes a superconductor. We rewrite the (\ref{EOMr1}) as the following:
\begin{eqnarray}
\frac{d}{dr}\Big[(-r^3+24\gamma\frac{r_{+}^4}{r})\phi^{\prime }\Big]%
+(1+8\gamma\frac{r_{+}^4}{r^4})\frac{\psi^{2}\phi}{rr_{+}^2f}=0 .
\label{pertubphi}
\end{eqnarray}
In the normal phase when the characteristic temperature of the system (the black hole temperature) is larger than the critical one, i.e. $T>T_c$ the scalar field is absent $\psi=0$. As the zeroth order approximation, the system has the following solution for the zeroth order gauge field $\phi_0$:
\begin{eqnarray}
\frac{d}{dr}\Big[(-r^3+24\gamma\frac{r_{+}^4}{r})\phi_0^{\prime }\Big]=0\label{pertubphi}.
\end{eqnarray}
To study the (in)stability of the system, we perturb $\phi(r)
= \phi_0(r) + \delta \phi$., where the (\ref{pertubphi}) gives us the following zeroth order equality:
\begin{eqnarray}
\frac{d}{dr}\Big[(-r^3+24\gamma\frac{r_{+}^4}{r})\delta\phi^{\prime }\Big] %
\geq 0,
\end{eqnarray}
It is clear for us that at the AdS boundary, 
\begin{eqnarray}
(-r^3+24\gamma\frac{r_{+}^4}{r})\phi_0^{\prime }\to 0,
\end{eqnarray}
consequently,we conclude that:
\begin{eqnarray}
(-r^3+24\gamma\frac{r_{+}^4}{r}) \delta\phi^{\prime }\to 0.
\end{eqnarray}
At AdS boundary,using this fact that always  $\delta\phi |_{r_+}= 0\Longrightarrow \delta
\phi^{\prime }\leq 0 $ so we conclude that:
\begin{eqnarray}
\phi(r) \leq \phi_0(r). \
\end{eqnarray}
It implies that there exists a physical state which it remains always below the zeroth order (normal) phase. 
But to have a complete proof, we rewrite the second field equation of the scalar field in the following form:
\begin{eqnarray}
\frac{d}{dr}\Big[(-r^3+24\gamma\frac{r_{+}^4}{r})\psi^{\prime }\Big]%
-(1+8\gamma\frac{r_{+}^4}{r^4})\frac{\psi^{2}\phi}{rr_{+}^2f^2}=0 .
\label{pertubpsi}
\end{eqnarray}
We define an auxiliary  field $X=r\psi$. In terms of this new defined field, the boundary condition $%
\psi^{\prime }(r_+)=0$ implies that $%
X^{\prime }_+ = X_+/r_+$. As we know at the AdS asymptotic limit, $rfX^{\prime }\to
0$. So to show that the system has a unique condensed phase, we need to find a critical point of the function $X(r)$ such that it satisfies the following equation: 
\begin{eqnarray}
\Big(X^{\prime }(r_{_T})=0\Big)\wedge \Big(X^{\prime \prime }<0\Big)\wedge \Big(X>0\Big).
\end{eqnarray}
Which is always possible to find it as the following form:
\begin{eqnarray}
X(r)=X(r_{T})+\frac{1}{2}X^{\prime \prime }(r_{T})(r-r_{T})^2,\ \ X^{\prime \prime }(r_{T})<0.
\end{eqnarray}

We have been qualitatively verified that the p-wave Weyl
superconductor state  exists .
%%%%%%%%%%%%%%%%%%%%%%%%%%%%%%%%%%%%%%%%%%%%%%%%%%%%%%%%%%%%%%%%%%%%%%%%%%%%

%\subsection{Critical temperature and condensation values by matching method}

%%%%%%%%%%%%%%%%%%%%%%%%%%%%%%%%%%%%%%%%%%%%%%%%%%%%%%%%%%%%%%%%%%%%%%%%%%%%%%
%\textbf{completed}
The apprpriate boundary conditions are given by the following:
\begin{eqnarray}
\phi(1)=0\,,\hspace{1cm}\psi^\prime(1)=0\,,  \label{regularity}
\end{eqnarray}
We use the matching method. Let us to start by writring the series solutions in the vicinity of the boundary points $z=1$ and $z=0$

%%%%%%%%%%%%%%%%%%%%%%%%%%%%%%%%%%%%%%%%%%%

%\subsection{Solution near the horizon}

%%%%%%%%%%%%%%%%%%%%%%%%%%%%%%%%%%%%%%%%%%%
The series solutions for $\{\phi,\psi\}$ are valid in the vicinity of the black hole horizon  $z=1$ as follows:
\begin{eqnarray}
\phi(z)&=&\phi(1)-\phi^\prime(1)(1-z)+\frac{1}{2}\phi^{\prime%
\prime}(1)(1-z)^2 +\cdots ,  \label{series1} \\
\psi(z)&=&\psi(1)-\psi^\prime(1)(1-z)+\frac{1}{2}\psi^{\prime%
\prime}(1)(1-z)^2 +\cdots .  \label{series2}
\end{eqnarray}
We put $a=-\phi^\prime(1)<0$ and $b=\psi(1)>0$ . Using the (\ref{EOMz1}) we are able to find the coefficent:
\begin{eqnarray}
\phi^{\prime \prime }(1)&=&-a\frac{1+72\gamma}{1-24\gamma}+\frac{1+8\gamma}{%
1-24\gamma} \frac{ab^2}{4r_{+}^2}.
\end{eqnarray}
Thus we can write  (\ref{series1}) for the gauge field:
\begin{eqnarray}
\phi(z)=a\Big[(1-z)-\frac{(1-z)^2}{2(1-24\gamma)}\Big(1+72\gamma-\frac{b^2}{%
4r_{+}^2}(1+8\gamma)\Big)\Big] .  \label{seriesphi}
\end{eqnarray}
and  similarly for $\psi$, using  (\ref{EOMz2}) we obtain:
\begin{eqnarray}
\psi^{\prime \prime }(1)&=&-\frac{1}{32r_{+}^2}\Big(\frac{1+8\gamma}{%
1-8\gamma}\Big)a^2b.
\end{eqnarray}
appropriatly the following series solution is written for the scalar field: (\ref{series2})
\begin{eqnarray}
\psi(z)&=&b\Big[1-\Big(\frac{1+8\gamma}{1-8\gamma}\Big)\frac{(1-z)^2}{%
64r_{+}^2}a^2\Big] .  \label{seriespsi}
\end{eqnarray}
%%%%%%%%%%%%%%%%%%%%%%%%%%%
%\subsection{Solution near the asymptotic AdS region}

We need also to provide the asymptotic solutions near the AdS boundary. For the EOMs given in  (\ref{EOMz1}) and (\ref{EOMz2}), and in the asymptotic limit we have: region we have
\begin{eqnarray}
\phi(z)&=&\mu-\frac{\rho}{r_{+}^2}z^2 ,  \label{ads1} \\
\psi(z)&=&\frac{<\mathcal{O_{+}>}}{r_{+}^\Delta}z^\Delta .  \label{ads2}
\end{eqnarray}
We adopted a suitable renormalization $\mathcal{O}_{-}>=0$. To have a better overview, we are remembering to the mind that the scalar field has the following exact solution near the AdS boundary:
\begin{eqnarray}
\psi(r)\rightarrow r^{-(\Delta_{+}+2)}+ r^{-\Delta_{+}}<\mathcal{O_{+}>\,.}
\end{eqnarray}
The unique duality with the CFT operators is clearly understood now from this expansion, since according to the complex variable theory, the coefficents of the series are determined uniquely as countour integrals over an arbitrary contour. Here the order of series is given by the power  $\Delta_{+}$. We are free to  omit the terms with the coefficents as  $<\mathcal{O_{-}}>$. It is simply done by putting $\mathcal{O_{-}}=0$. 

%%%%%%%%%%%%%%%%%%%%%%%%%%%%
%\subsection{Matching and phase transition}

Now, we are going to smoothly match the solutions given by (\ref{seriesphi}) and (\ref{seriespsi}) with (\ref{ads1}) and (\ref{ads2}) at an intermediate point $z=z_m$. To be smoothly connected we must check:
\begin{eqnarray}
\phi^{z=0}|_{z_m}=\phi^{z=1}|_{z_m},\ \ \phi^{\prime
z=0}|_{z_m}=\phi^{\prime z=1}|_{z_m} \\
\psi^{z=0}|_{z_m}=\psi^{z=1}|_{z_m},\ \ \psi^{\prime
z=0}|_{z_m}=\psi^{\prime z=1}|_{z_m}
\end{eqnarray}
If we substitute the functions in the above mentioned continuity conditions we obtain:
\begin{eqnarray}
&&\mu-{\frac {\rho\,{z_{{m}}}^{2}}{{\pi }^{2}{T}^{2}}}=a \left( 1-z_{{m} }-%
\frac{1}{2}\,{\frac { \left( 1-z_{{m}} \right) ^{2} \left( 1+72\,\gamma-%
\frac{1}{4r_{+}^2} \,{b}^{2} \left( 1+8\,\gamma \right) \right) }{%
1-24\,\gamma}} \right),  \label{c:phi} \\
&&-\,{\frac {2\rho\,z_{{m}}}{{\pi }^{2}{T}^{2}}}=a \left( -1+{\frac { \left(
1-z_{{m}} \right) \left( 1+72\,\gamma-\frac{1}{4r_{+}^2}\,{b}^{2} \left( 1+8
\,\gamma \right) \right) }{1-24\,\gamma}} \right),  \label{c:dphi} \\
&&\frac {{z_{{m}}}^{\Delta}<\mathcal{O_{+}>}}{(\pi T)^{\Delta}}=b \left( 1-{%
\ \frac {1}{64r_{+}^2}}\,{\frac { \left( 1+8\,\gamma \right) {a}^{2} \left(
1-z _{{m}} \right) ^{2}}{1-8\,\gamma}} \right),  \label{c:psi} \\
&&\,{\frac {\Delta z_{m}^{\Delta-1}<\mathcal{O_{+}>}}{(\pi T)^{\Delta}}}=%
\frac{1}{32r_{+}^2}\,{\ \frac {b \left( 1+8\,\gamma \right) {a}^{2} \left(
1-z_{{m}} \right) } {1-8\,\gamma}} .  \label{c:dpsi}
\end{eqnarray}
Since we've$r_+=\pi T$, if we eliminate $a b^2$ from (\ref{c:phi}) and
(\ref{c:dphi}) we find:
\begin{eqnarray}
\mu&=&\frac{1}{2}\,{\frac {2\,\rho\,z_{{m}}-az_{{m}}{\pi }^{2}{T}^{2}+a{\pi }%
^ {2}{T}^{2}}{{\pi }^{2}{T}^{2}}}, \\
b&=&\sqrt {-{\frac {-8\,\rho\,z_{{m}}+192\,\rho\,z_{{m}}\gamma-384\,a
\gamma\,{\pi }^{2}{T}^{2}+4\,az_{{m}}{\pi }^{2}{T}^{2}+288\,az_{{m}} \gamma\,%
{\pi }^{2}{T}^{2}}{a(1+8\,\gamma-z_{{m}}-8\,z_{{m}}\gamma)}}} .  \label{b}
\end{eqnarray}
Similarly we eliminate  $a^2b$  between (\ref{c:psi}) and (\ref{c:dpsi}) . By this simple manipulation 
we find the  non-vanishing $b$, and later we can eliminate $<\mathcal{O_{+}>}$ between (\ref{c:psi})
, (\ref{c:dpsi}) to lead us:
\begin{eqnarray}
a&=&8\pi T{\frac { \,\sqrt { \left(1- z_{{m}} \right) \left( -\Delta-2\, z_{{%
m}}+z_{{m}}\Delta \right) \left( 1+8\,\gamma \right) \Delta\, \left(
-1+8\,\gamma \right) }}{ \left( z_{{m}}-1 \right) \left( - \Delta-2\,z_{{m}%
}+z_{{m}}\Delta \right) \left( 1+8\,\gamma \right) }}.
\end{eqnarray}

If we replace $a$, from (\ref{b}) we discover that:

\begin{eqnarray}
b &=&\sqrt{36\left( \left( \gamma +{\frac{1}{72}}\right) z_{{m}}-\frac{4}{3}%
\,\gamma \right) {\pi }^{3}\sqrt{(1-z_{{m}})\left( -\Delta -2\,z_{{m}}+z_{{m}%
}\Delta \right) \left( 1+8\,\gamma \right) \Delta \,\left( -1+8\,\gamma
\right) }}  \notag \\
&&\times \sqrt{{\frac{{T}^{3}-T_{c}^{3}}{\left( \frac{1}{8}+\gamma \right)
\left( -1+z_{{m}}\right) T\sqrt{\left( 1-z_{{m}}\right) \left( 1+8\,\gamma
\right) \left( -1+8\,\gamma \right) \Delta \,\left( -2\,z_{{m}}+\Delta \,z_{{%
m}}-\Delta \right) }\pi }}}.
\end{eqnarray}%
where we define the critical temperature $T_{c}$ as the following:
\begin{equation}
T_{c}=\alpha (\gamma |z_{m}|\Delta )\rho ^{\frac{1}{3}}  \label{Tc2}
\end{equation}%
here ,
\begin{equation*}
\alpha (\gamma |z_{m}|\Delta )=\frac{1}{2}\,\sqrt[3]{2(1+8\,\gamma
)(-1+24\,\gamma )}\times \sqrt[3]{\frac{z_{m}}{z_{m}(1+72\gamma )-96\gamma }%
\sqrt{\frac{(1-z_{m})(-\Delta +z_{m}(\Delta -2))}{\Delta }}}
\end{equation*}%
Because of the acceptable range of Weyl's coupling, i.e. $-\frac{1}{16}<\gamma <\frac{1}{24}$ and since the temperature (critical) must be positive, we should consider the following cases separately:

\begin{itemize}
\item Case $A:\ -\frac{1}{16}<\gamma<-\frac{1}{8}, \gamma>\frac{z_m}{96-72z_m%
}$,

\item Case $B:\ -\frac{1}{8}<\gamma<\frac{1}{24}, \gamma<\frac{z_m}{96-72z_m}
$.
\end{itemize}

It is adequate to mention here that in both cases, the conformal dimension reads  $\Delta>-\frac{2z_m}{1-z_m}$. Now we can also write the expression for 
 $\langle\mathcal{O}_+\rangle$ as the following:
\begin{eqnarray}
\langle\mathcal{O}_{+}\rangle =\beta(\gamma|z_m|\Delta)T_c^{\frac{3}{2}%
}T^{\Delta-\frac{1}{2}}\sqrt{1-(\frac{T}{T_c})^3},
\end{eqnarray}
where
\begin{eqnarray}
\beta(\gamma|z_m|\Delta)&=&\frac{\sqrt {36\left( \left( \gamma+{\frac {1}{72}%
} \right) z_{{m}}-\frac{4}{3}\,\gamma \right) 
}}{{z_{{m}}}^{\Delta-1} ( \frac{1}{2}\,\Delta+z_{{m}} ( 1-\frac{1}{2}%
\,\Delta ))}   \frac{\pi^{\Delta+1/2}}{\sqrt{\mathcal{A}} }.
\end{eqnarray}
here $\mathcal{A}=\sqrt {( 1-z_{{m}} )
( -\Delta-2 \,z_{{m}}+z_{{m}}\Delta)( 1+8\,\gamma) \Delta\, ( -1+8\,\gamma) }$.
Using the definition of $T_c$, we deduce that in our model $\langle \mathcal{O}_{+}\rangle $ to be vanished at the critical point $T=T_ {c} $, and there exists a unique superconducting phase $TT_ {c} $. The result for $\langle \mathcal{O}_{+}\rangle $ to be confirmed by the mean field theory result $ \langle \mathcal{O}_{+}\rangle \propto (1-T/T_ {c}) ^ {1/2} $ . Also, the value of $T_ {c} $given by (\ref{Tc2}) is considered in greater agreement with the numerical results as $T_{c}\propto \sqrt[3]{\rho } $ \cite{Momeni:2012ab}.

%%%%%%%%%%%%%%%%%%%%%%%%%%
\section{Conclusion}
%%%%%%%%%%%%%%%%%%%%%%%%%%%%
In this paper, we review the basic properties of s, p-wave holographic superconductors with the Weyl correction term. Such corrections are motivated by one-loop quantum corrections to the Einstein gravity. The main focus was on analytical properties. Matching method and variations approach have been used to derive exact analytic results for critical temperature and condensation of different conformal operators. We also has been investigated a type of generalized models with St\'{u}ckelberg 's and analytically well established the critical parameters of this model. In connection with this mini review of higher order Weyl corrections ,it is of great importance to mention, that however such a higher order corrected model may be convenient in investigating effects to some kind of superconductors, yet it is in vain to attempt a clear comparison of the condensation, founded upon Gauss-Bonnet correction with our case. In Gauss-Bonnet holographic superconductors, the coupling constant makes the condensation harder. The critical temperature decreased significantly, so the condensation becomes more colder. In contrast to this case, our Weyl corrected model makes condensation less  harder. The reason may be these two types of the higher order cirvature corrections have the same order , indeed $GB\sim W^2$ (W is Weyl tensor). So,  ; notwithstanding that those marked Weyl parts or features very obviously seem weaker adapted to afford the basis for a regular system of holographic superconductor than any other higher order corrected system, which the Riemannian tensor present. Finally we mention here that it might also be interesting in future to investigate holographic superconductors in context with RS braneworld with GB in the bulk \cite{Dufaux:2004qs},\cite{Tsujikawa:2004dm}.

%%%%%%%%%%%%%%%%%%%%%%%%%%%%%%%%%%%%%%%%%%%%%%%%%%%%%%%%%%%%%%%%%%%%%%%%%%%%%
%\begin{acknowledgments} 
%We gratefully acknowledge all of our co-authors which shared their results with us.
%M. Sami and S. D. Odintsov for useful comments.
%\end{acknowledgments}

%%%%%%%%%%%%%%%%%%%%%%%%%%%%%%%%%%%%%%%%%%%%%%%%%%%%%%%%%%%%%%%%%%%%%%%%%%%%%
%\clearpage
%\appendix

%%%%%%%%%%%%%%%%%%%%%%%%%%%%%

%%%%%%%%%%%%%%%%%%%%%%%%%%%%


\begin{thebibliography}{90}
%%%%%%%%%%%%%%%%%%%%%
%Introduction
%%%%%%%%%%%%%%%%%%%%%%

%\cite{Maldacena:1997re}
\bibitem{Maldacena:1997re}
  J.~M.~Maldacena,
  ``The large N limit of superconformal field theories and supergravity,''
  Adv.\ Theor.\ Math.\ Phys.\  {\bf 2}, 231 (1998)
  [Int.\ J.\ Theor.\ Phys.\  {\bf 38}, 1113 (1999)]
  [arXiv:hep-th/9711200].
  %%CITATION = IJTPB,38,1113;%%
%\cite{Gubser:1998bc}
\bibitem{Gubser:1998bc}
  S.~S.~Gubser, I.~R.~Klebanov and A.~M.~Polyakov,
  ``Gauge theory correlators from non-critical string theory,''
  Phys.\ Lett.\  B {\bf 428}, 105 (1998)
  [arXiv:hep-th/9802109].
  %%CITATION = PHLTA,B428,105;%%
%\cite{Witten:1998qj}
\bibitem{Witten:1998qj}
  E.~Witten,
  ``Anti-de Sitter space and holography,''
  Adv.\ Theor.\ Math.\ Phys.\  {\bf 2}, 253 (1998)
  [arXiv:hep-th/9802150].
  %%CITATION = 00203,2,253;%%


%%%%%%%%%%%%%%%%%%%review %%%%%%%%%%%%%%%

%\cite{Hartnoll:2009sz}
\bibitem{Hartnoll:2009sz}
  S.~A.~Hartnoll,
``Lectures on holographic methods for condensed matter physics,''
 Class.\ Quant.\ Grav.\  {\bf 26}, 224002 (2009)  [arXiv:0903.3246 [hep-th]].
 %%CITATION = ARXIV:0903.3246;%%

%\cite{Herzog:2009xv}
\bibitem{Herzog:2009xv}
  C.~P.~Herzog,
  ``Lectures on Holographic Superfluidity and Superconductivity,''
  J.\ Phys.\ A {\bf 42}, 343001 (2009)
  [arXiv:0904.1975 [hep-th]].
  %%CITATION = ARXIV:0904.1975;%%

%\cite{McGreevy:2009xe}
\bibitem{McGreevy:2009xe}
  J.~McGreevy,
  ``Holographic duality with a view toward many-body physics,''
  Adv.\ High Energy Phys.\  {\bf 2010}, 723105 (2010)
  [arXiv:0909.0518 [hep-th]].
  %%CITATION = ARXIV:0909.0518;%%
%%%%%%%%%%%%

%\cite{Horowitz:2010gk}
\bibitem{Horowitz:2010gk}
  G.~T.~Horowitz,
  ``Introduction to Holographic Superconductors,''  Lect.\ Notes Phys.\  {\bf 828}, 313 (2011)  [arXiv:1002.1722 [hep-th]].
   %%CITATION = ARXIV:1002.1722;%%  %245 citations counted in INSPIRE as of 02 Jan 2014




%\cite{Horowitz:2012ky}
\bibitem{Horowitz:2012ky}
  G.~T.~Horowitz, J.~E.~Santos and D.~Tong,
  ``Optical Conductivity with Holographic Lattices,''
  JHEP {\bf 1207}, 168 (2012)
  [arXiv:1204.0519 [hep-th]].
  %%CITATION = ARXIV:1204.0519;%%

%\cite{Horowitz:2013jaa}
\bibitem{Horowitz:2013jaa}
  G.~T.~Horowitz and J.~E.~Santos,
  ``General Relativity and the Cuprates,''
  arXiv:1302.6586 [hep-th].
  %%CITATION = ARXIV:1302.6586;%%

  %\cite{Ling:2013aya}
\bibitem{Ling:2013aya}
  Y.~Ling, C.~Niu, J.~-P.~Wu, Z.~-Y.~Xian and H.~-b.~Zhang,
  ``Holographic Fermionic Liquid with Lattices,''
  JHEP {\bf 1307}, 045 (2013)
  [arXiv:1304.2128 [hep-th]].
  %%CITATION = ARXIV:1304.2128;%%

%\cite{Donos:2011bh}
\bibitem{Donos:2011bh}
  A.~Donos and J.~P.~Gauntlett, ``Holographic striped phases,''  JHEP {\bf 1108}, 140 (2011)  [arXiv:1106.2004 [hep-th]].
%%CITATION = ARXIV:1106.2004;%%  %64 citations counted in INSPIRE as of 22 Jan 2014

%\cite{Donos:2013wia}
\bibitem{Donos:2013wia}
  A.~Donos,``Striped phases from holography,''  JHEP {\bf 1305}, 059 (2013)  [arXiv:1303.7211 [hep-th]].  
%%CITATION = ARXIV:1303.7211;%%  %13 citations counted in INSPIRE as of 23 Jan 2014



%\cite{Rozali:2013ama}
\bibitem{Rozali:2013ama}
  M.~Rozali, D.~Smyth, E.~Sorkin and J.~B.~Stang,
  ``Striped Order in AdS/CFT,''
  Phys.\ Rev.\ D {\bf 87}, 126007 (2013)
  [arXiv:1304.3130 [hep-th]].
  %%CITATION = ARXIV:1304.3130;%%

  %\cite{Cai:2013sua}
\bibitem{Cai:2013sua}
  R.~-G.~Cai, Y.~-Q.~Wang and H.~-Q.~Zhang,
  ``A holographic model of SQUID,''
  arXiv:1308.5088 [hep-th].
  %%CITATION = ARXIV:1308.5088;%%



%\cite{Murata:2010dx}
\bibitem{Murata:2010dx}
  K.~Murata, S.~Kinoshita and N.~Tanahashi,
  ``Non-equilibrium Condensation Process in a Holographic Superconductor,''
  JHEP {\bf 1007}, 050 (2010)
  [arXiv:1005.0633 [hep-th]].
  %%CITATION = ARXIV:1005.0633;%%

%\cite{Bhaseen:2012gg}
\bibitem{Bhaseen:2012gg}
  M.~J.~Bhaseen, J.~P.~Gauntlett, B.~D.~Simons, J.~Sonner and T.~Wiseman,
  ``Holographic Superfluids and the Dynamics of Symmetry Breaking,''
  Phys.\ Rev.\ Lett.\  {\bf 110}, 015301 (2013)
  [arXiv:1207.4194 [hep-th]].
  %%CITATION = ARXIV:1207.4194;%%

%\cite{Adams:2012pj}
\bibitem{Adams:2012pj}
  A.~Adams, P.~M.~Chesler and H.~Liu,
  ``Holographic Vortex Liquids and Superfluid Turbulence,''
  arXiv:1212.0281 [hep-th].
  %%CITATION = ARXIV:1212.0281;%%

%\cite{Garcia-Garcia:2013rha}
\bibitem{Garcia-Garcia:2013rha}
  A.~M.~Garc��a-Garc��a, H.~B.~Zeng and H.~Q.~Zhang,
  ``A thermal quench induces spatial inhomogeneities in a holographic superconductor,''
  arXiv:1308.5398 [hep-th].
  %%CITATION = ARXIV:1308.5398;%%

  %\cite{Chesler:2013lia}
\bibitem{Chesler:2013lia}
  P.~M.~Chesler and L.~G.~Yaffe,
  ``Numerical solution of gravitational dynamics in asymptotically anti-de Sitter spacetimes,''
  arXiv:1309.1439 [hep-th].
  %%CITATION = ARXIV:1309.1439;%%




%\cite{Hartnoll:2008vx}
\bibitem{Hartnoll:2008vx}
S.~A.~Hartnoll, C.~P.~Herzog and G.~T.~Horowitz,
``Building a Holographic Superconductor,''
Phys.\ Rev.\ Lett.\  {\bf 101}, 031601 (2008)
[arXiv:0803.3295 [hep-th]].
%%CITATION = ARXIV:0803.3295;%%

%\cite{Hartnoll:2008kx}
\bibitem{Hartnoll:2008kx}
  S.~A.~Hartnoll, C.~P.~Herzog and G.~T.~Horowitz,
  ``Holographic Superconductors,''  JHEP {\bf 0812}, 015 (2008)  [arXiv:0810.1563 [hep-th]].
   %%CITATION = ARXIV:0810.1563;%%


%\cite{Nishioka:2009zj}
\bibitem{Nishioka:2009zj}
  T.~Nishioka, S.~Ryu and T.~Takayanagi,
  ``Holographic Superconductor/Insulator Transition at Zero Temperature,''
  JHEP {\bf 1003}, 131 (2010)
  [arXiv:0911.0962 [hep-th]].
  %%CITATION = ARXIV:0911.0962;%%

%\cite{Witten:1998zw}
\bibitem{Witten:1998zw}
  E.~Witten,
  ``Anti-de Sitter space, thermal phase transition, and confinement in gauge theories,''
  Adv.\ Theor.\ Math.\ Phys.\  {\bf 2}, 505 (1998)
  [hep-th/9803131].
  %%CITATION = HEP-TH/9803131;%%


%\cite{Horowitz:2010jq}
\bibitem{Horowitz:2010jq}
  G.~T.~Horowitz and B.~Way,
  ``Complete Phase Diagrams for a Holographic Superconductor/Insulator System,''
  JHEP {\bf 1011}, 011 (2010)
  [arXiv:1007.3714 [hep-th]].
  %%CITATION = ARXIV:1007.3714;%%



%\cite{Gubser:2008wv}
\bibitem{Gubser:2008wv}
  S.~S.~Gubser and S.~S.~Pufu,
  ``The Gravity dual of a p-wave superconductor,''  JHEP {\bf 0811}, 033 (2008)  [arXiv:0805.2960 [hep-th]].
   %%CITATION = ARXIV:0805.2960;%%



%\cite{Aprile:2010ge}
\bibitem{Aprile:2010ge}
  F.~Aprile, D.~Rodriguez-Gomez and J.~G.~Russo,
  ``p-wave Holographic Superconductors and five-dimensional gauged Supergravity,''
  JHEP {\bf 1101}, 056 (2011)
  [arXiv:1011.2172 [hep-th]].
  %%CITATION = ARXIV:1011.2172;%%
%%%%%%%%%%%%%%%%%%%%%%
%GB
\bibitem{Pan:2009xa} 
  Q.~Pan, B.~Wang, E.~Papantonopoulos, J.~Oliveira and A.~B.~Pavan,
  ``Holographic Superconductors with various condensates in Einstein-Gauss-Bonnet gravity,''
  Phys.\ Rev.\ D {\bf 81}, 106007 (2010)
  [arXiv:0912.2475 [hep-th]].
%\cite{Brihaye:2010mr}
\bibitem{Brihaye:2010mr} 
  Y.~Brihaye and B.~Hartmann,
  ``Holographic Superconductors in 3+1 dimensions away from the probe limit,''
  Phys.\ Rev.\ D {\bf 81}, 126008 (2010)
  [arXiv:1003.5130 [hep-th]].
%\cite{Pan:2010at}
\bibitem{Pan:2010at} 
  Q.~Pan and B.~Wang,
  ``General holographic superconductor models with Gauss-Bonnet corrections,''
  Phys.\ Lett.\ B {\bf 693}, 159 (2010)
  [arXiv:1005.4743 [hep-th]].
%\cite{Cai:2010cv}
\bibitem{Cai:2010cv} 
  R.~G.~Cai, Z.~Y.~Nie and H.~Q.~Zhang,
  ``Holographic p-wave superconductors from Gauss-Bonnet gravity,''
  Phys.\ Rev.\ D {\bf 82}, 066007 (2010)
  [arXiv:1007.3321 [hep-th]].
%\cite{Kuang:2010jc}
\bibitem{Kuang:2010jc} 
  X.~M.~Kuang, W.~J.~Li and Y.~Ling,
  ``Holographic Superconductors in Quasi-topological Gravity,''
  JHEP {\bf 1012}, 069 (2010)
  [arXiv:1008.4066 [hep-th]].
%\cite{Barclay:2010up}
\bibitem{Barclay:2010up} 
  L.~Barclay, R.~Gregory, S.~Kanno and P.~Sutcliffe,
  ``Gauss-Bonnet Holographic Superconductors,''
  JHEP {\bf 1012}, 029 (2010)
  [arXiv:1009.1991 [hep-th]].
%\cite{Jing:2010cx}
\bibitem{Jing:2010cx} 
  J.~Jing, L.~Wang, Q.~Pan and S.~Chen,
  ``Holographic Superconductors in Gauss-Bonnet gravity with Born-Infeld electrodynamics,''
  Phys.\ Rev.\ D {\bf 83}, 066010 (2011)
  [arXiv:1012.0644 [gr-qc]].
%\cite{Gregory:2010yr}
\bibitem{Gregory:2010yr} 
  R.~Gregory,
  ``Holographic Superconductivity with Gauss-Bonnet gravity,''
  J.\ Phys.\ Conf.\ Ser.\  {\bf 283}, 012016 (2011)
  [arXiv:1012.1558 [hep-th]].
%\cite{Barclay:2010nm}
\bibitem{Barclay:2010nm} 
  L.~Barclay,
  ``The Rich Structure of Gauss-Bonnet Holographic Superconductors,''
  JHEP {\bf 1110}, 044 (2011)
  [arXiv:1012.3074 [hep-th]].
%\cite{Cai:2010zm}
\bibitem{Cai:2010zm} 
  R.~G.~Cai, Z.~Y.~Nie and H.~Q.~Zhang,
  ``Holographic Phase Transitions of P-wave Superconductors in Gauss-Bonnet Gravity with Back-reaction,''
  Phys.\ Rev.\ D {\bf 83}, 066013 (2011)
  [arXiv:1012.5559 [hep-th]].
%\cite{Kanno:2011cs}
\bibitem{Kanno:2011cs} 
  S.~Kanno,
  ``A Note on Gauss-Bonnet Holographic Superconductors,''
  Class.\ Quant.\ Grav.\  {\bf 28}, 127001 (2011)
  [arXiv:1103.5022 [hep-th]].
%\cite{Jing:2012dj}
\bibitem{Jing:2012dj} 
  J.~Jing, Q.~Pan and S.~Chen,
  ``Holographic Superconductor/Insulator Transition with logarithmic electromagnetic field in Gauss-Bonnet gravity,''
  Phys.\ Lett.\ B {\bf 716}, 385 (2012)
  [arXiv:1209.0893 [hep-th]].
%\cite{Roychowdhury:2013aua}
\bibitem{Roychowdhury:2013aua} 
  D.~Roychowdhury,
  ``Holographic droplets in p-wave insulator/superconductor transition,''
  JHEP {\bf 1305}, 162 (2013)
  [arXiv:1304.6171 [hep-th]].
%\cite{Yao:2013sha}
\bibitem{Yao:2013sha} 
  W.~Yao and J.~Jing,
  ``Analytical study on holographic superconductors for Born-Infeld electrodynamics in Gauss-Bonnet gravity with backreactions,''
  JHEP {\bf 1305}, 101 (2013)
  [arXiv:1306.0064 [gr-qc]].
%\cite{Dey:2013qoa}
\bibitem{Dey:2013qoa} 
  S.~Dey and A.~Lala,
  ``Holographic $s$-wave condensation and Meissner-like effect in Gauss-Bonnet gravity with various non-linear corrections,''
  arXiv:1306.5137 [hep-th].
%\cite{Hartmann:2013nla}
\bibitem{Hartmann:2013nla} 
  B.~Hartmann,
  ``Stability of black holes and solitons in Anti-de Sitter space-time,''
  Nucl.\ Phys.\ Proc.\ Suppl.\  {\bf 251-252}, 45 (2014)
  [arXiv:1310.0300 [gr-qc]].
%\cite{Wu:2014lta}
\bibitem{Wu:2014lta} 
  Y.~B.~Wu, J.~W.~Lu, Y.~Y.~Jin, J.~B.~Lu, X.~Zhang, S.~Y.~Wu and C.~Wang,
  ``Magnetic-field effects on $p$-wave phase transition in Gauss-Bonnet gravity,''
  Int.\ J.\ Mod.\ Phys.\ A {\bf 29}, no. 20, 1450094 (2014)
  [arXiv:1405.2499 [hep-th]].
  %%CITATION = ARXIV:1405.2499





%%%%%%%%%%%%%%%%%%%%%%
%s-wave
%%%%%%%%%%%%%%%%%%%%%%
\bibitem{Wu:2010vr} 
  J.~P.~Wu, Y.~Cao, X.~M.~Kuang and W.~J.~Li,
  ``The 3+1 holographic superconductor with Weyl corrections,''
  Phys.\ Lett.\ B {\bf 697}, 153 (2011)
  [arXiv:1010.1929 [hep-th]].
\bibitem{qc} I. T. Drummond and S. J. Hathrell, Phys. Rev. D 22, 343 (1980).
%\cite{Breitenlohner:1982bm}
\bibitem{Breitenlohner:1982bm} 
  P.~Breitenlohner and D.~Z.~Freedman,
  ``Positive Energy in anti-De Sitter Backgrounds and Gauged Extended Supergravity,''
  Phys.\ Lett.\ B {\bf 115}, 197 (1982).



%%%%%%%%%%%%%%%%%%
%Analytical scheme
%%%%%%%%%%%%%
%\cite{Gregory:2009fj}
\bibitem{Gregory:2009fj} 
  R.~Gregory, S.~Kanno and J.~Soda,
  ``Holographic Superconductors with Higher Curvature Corrections,''
  JHEP {\bf 0910}, 010 (2009)
  [arXiv:0907.3203 [hep-th]].

%\cite{Bai:2014poa}
\bibitem{Bai:2014poa} 
  X.~Bai, B.~H.~Lee, M.~Park and K.~Sunly,
  ``Dynamical Condensation in a Holographic Superconductor Model with Anisotropy,''
  JHEP {\bf 1409}, 054 (2014)
  [arXiv:1405.1806 [hep-th]].
%\cite{Momeni:2014rla}
\bibitem{Momeni:2014rla} 
  D.~Momeni and R.~Myrzakulov,
  ``Universality of critical magnetic field in holographic superconductor,''
  arXiv:1401.3658 [hep-th].
%\cite{Banerjee:2013maa}
\bibitem{Banerjee:2013maa} 
  N.~Banerjee, S.~Dutta and D.~Roychowdhury,
  ``Chern-Simons Superconductor,''
  arXiv:1311.7640 [hep-th].
  %%CITATION = ARXIV:1311.7
%\cite{Zhao:2013pva}
\bibitem{Zhao:2013pva} 
  Z.~Zhao, Q.~Pan and J.~Jing,
  ``Notes on analytical study of holographic superconductors with Lifshitz scaling in external magnetic field,''
  Phys.\ Lett.\ B {\bf 735}, 438 (2014)
  [arXiv:1311.6260 [hep-th]].

%\cite{Gangopadhyay:2013qza}
\bibitem{Gangopadhyay:2013qza} 
  S.~Gangopadhyay,
  ``Holographic superconductors in Born-Infeld electrodynamics and external magnetic field,''
  Mod.\ Phys.\ Lett.\ A {\bf 29}, 1450088 (2014)
  [arXiv:1311.4416 [hep-th]].

%\cite{Huang:2013sca}
\bibitem{Huang:2013sca} 
  W.~H.~Huang,
  ``Analytic Study of First-Order Phase Transition in Holographic Superconductor and Superfluid,''
  Int.\ J.\ Mod.\ Phys.\ A {\bf 28}, 1350140 (2013)
  [arXiv:1307.5614 [hep-th]].
%\cite{Momeni:2013via}
\bibitem{Momeni:2013via} 
  D.~Momeni, M.~Raza and R.~Myrzakulov,
  ``Construction of a Holographic Superconductor in F(R) Gravity,''
  Eur.\ Phys.\ J.\ Plus {\bf 129}, 30 (2014)
  [arXiv:1307.2497 [hep-th]].
%\cite{Li:2013glh}
\bibitem{Li:2013glh} 
  H.~F.~Li,
  ``Further studies on holographic insulator/superconductor phase transitions from Sturm-Liouville eigenvalue problems,''
  JHEP {\bf 1307}, 135 (2013)
  [arXiv:1306.3071 [hep-th]].
%\cite{Momeni:2013eva}
\bibitem{Momeni:2013eva} 
  D.~Momeni, M.~Raza and R.~Myrzakulov,
  ``More on Superconductors via Gauge/Gravity Duality with Nonlinear Maxwell Field,''
  J.\ Grav.\  {\bf 2013}, 782512 (2013)
  [arXiv:1305.3541 [physics.gen-ph]].
%\cite{Cui:2013uha}
\bibitem{Cui:2013uha}
  S.~l.~Cui and Z.~Xue,
  ``Critical magnetic field in a holographic superconductor in Gauss-Bonnet gravity with Born-Infeld electrodynamics,''
  Phys.\ Rev.\ D {\bf 88} (2013) 10,  107501
  [arXiv:1306.2013 [hep-th]].
%\cite{Momeni:2013via}
\bibitem{Momeni:2013via} 
  D.~Momeni, M.~Raza and R.~Myrzakulov,
  ``Construction of a Holographic Superconductor in F(R) Gravity,''
  Eur.\ Phys.\ J.\ Plus {\bf 129}, 30 (2014)
  [arXiv:1307.2497 [hep-th]].
%\cite{Ge:2012vp}
\bibitem{Ge:2012vp}
  X.~H.~Ge, S.~F.~Tu and B.~Wang,
  ``d-Wave holographic superconductors with backreaction in external magnetic fields,''
  JHEP {\bf 1209} (2012) 088
  [arXiv:1209.4272 [hep-th]].
%\cite{Momeni:2012tw}
\bibitem{Momeni:2012tw} 
  D.~Momeni, R.~Myrzakulov, L.~Sebastiani and M.~R.~Setare,
  ``Analytical holographic superconductors in $AdS_N$ topological Lifshitz black holes,''
  arXiv:1210.7965 [hep-th].
%\cite{Gangopadhyay:2012gx}
\bibitem{Gangopadhyay:2012gx} 
  S.~Gangopadhyay and D.~Roychowdhury,
  ``Analytic study of properties of holographic p-wave superconductors,''
  JHEP {\bf 1208}, 104 (2012)
  [arXiv:1207.5605 [hep-th]].
%\cite{Ge:2011cw}
\bibitem{Ge:2011cw} 
  X.~H.~Ge and H.~Q.~Leng,
  ``Analytical calculation on critical magnetic field in holographic superconductors with backreaction,''
  Prog.\ Theor.\ Phys.\  {\bf 128}, 1211 (2012)
  [arXiv:1105.4333 [hep-th]].
%\cite{Momeni:2010jf}
\bibitem{Momeni:2010jf} 
  D.~Momeni, M.~R.~Setare and N.~Majd,
  ``Holographic superconductors in a model of non-relativistic gravity,''
  JHEP {\bf 1105}, 118 (2011)
  [arXiv:1003.0376 [hep-th]].

%\cite{Cai:2011ky}
\bibitem{Cai:2011ky}
  R.~G.~Cai, H.~F.~Li and H.~Q.~Zhang,
  ``Analytical Studies on Holographic Insulator/Superconductor Phase Transitions,''
  Phys.\ Rev.\ D {\bf 83} (2011) 126007
  [arXiv:1103.5568 [hep-th]].
%\cite{Setare:2011ip}
\bibitem{Setare:2011ip} 
  M.~R.~Setare and D.~Momeni,
  ``Gauss-Bonnet holographic superconductors with magnetic field,''
  Europhys.\ Lett.\  {\bf 96}, 60006 (2011)
  [arXiv:1106.1025 [physics.gen-ph]].
%\cite{Basu:2011np}
\bibitem{Basu:2011np} 
  P.~Basu,
  ``Low temperature properties of holographic condensates,''
  JHEP {\bf 1103}, 142 (2011)
  [arXiv:1101.0215 [hep-th]].
  %%CITATION = ARXIV:1101.021
%%%%%%%%%%%%%%%%%%%%%%%%%
%SL approach

\bibitem{Li:2011xja} 
  H.~F.~Li, R.~G.~Cai and H.~Q.~Zhang,
  ``Analytical Studies on Holographic Superconductors in Gauss-Bonnet Gravity,''
  JHEP {\bf 1104}, 028 (2011)
  [arXiv:1103.2833 [hep-th]].
%\cite{Pan:2011ah}
\bibitem{Pan:2011ah} 
  Q.~Pan, J.~Jing and B.~Wang,
  ``Analytical investigation of the phase transition between holographic insulator and superconductor in Gauss-Bonnet gravity,''
  JHEP {\bf 1111}, 088 (2011)
  [arXiv:1105.6153 [gr-qc]].
  %%CITATION = ARXIV:1105.61
%\cite{Jing:2011vz}
\bibitem{Jing:2011vz} 
  J.~Jing, Q.~Pan and S.~Chen,
  ``Holographic Superconductors with Power-Maxwell field,''
  JHEP {\bf 1111}, 045 (2011)
  [arXiv:1106.5181 [hep-th]].
%\cite{Momeni:2011iw}
\bibitem{Momeni:2011iw} 
  D.~Momeni, E.~Nakano, M.~R.~Setare and W.~Y.~Wen,
  ``Analytical study of critical magnetic field in a holographic superconductor,''
  Int.\ J.\ Mod.\ Phys.\ A {\bf 28}, 1350024 (2013)
  [arXiv:1108.4340 [hep-th]].
%\cite{Albrecht:2011xk}
\bibitem{Albrecht:2011xk} 
  D.~Albrecht, J.~Erlich and R.~J.~Wilcox,
  ``Nonlinear Boundary Dynamics and Chiral Symmetry in Holographic QCD,''
  Phys.\ Rev.\ D {\bf 85}, 114012 (2012)
  [arXiv:1112.5643 [hep-ph]].
%\cite{Gangopadhyay:2012am}
\bibitem{Gangopadhyay:2012am} 
  S.~Gangopadhyay and D.~Roychowdhury,
  ``Analytic study of properties of holographic superconductors in Born-Infeld electrodynamics,''
  JHEP {\bf 1205}, 002 (2012)
  [arXiv:1201.6520 [hep-th]].
%\cite{Gangopadhyay:2012np}
\bibitem{Gangopadhyay:2012np} 
  S.~Gangopadhyay and D.~Roychowdhury,
  ``Analytic study of Gauss-Bonnet holographic superconductors in Born-Infeld electrodynamics,''
  JHEP {\bf 1205}, 156 (2012)
  [arXiv:1204.0673 [hep-th]].
%\cite{Pan:2012jf}
\bibitem{Pan:2012jf} 
  Q.~Pan, J.~Jing, B.~Wang and S.~Chen,
  ``Analytical study on holographic superconductors with backreactions,''
  JHEP {\bf 1206}, 087 (2012)
  [arXiv:1205.3543 [hep-th]].
%\cite{Gangopadhyay:2012gx}
\bibitem{Gangopadhyay:2012gx} 
  S.~Gangopadhyay and D.~Roychowdhury,
  ``Analytic study of properties of holographic p-wave superconductors,''
  JHEP {\bf 1208}, 104 (2012)
  [arXiv:1207.5605 [hep-th]].
%\cite{Banerjee:2012vk}
\bibitem{Banerjee:2012vk} 
  R.~Banerjee, S.~Gangopadhyay, D.~Roychowdhury and A.~Lala,
  ``Holographic s-wave condensate with non-linear electrodynamics: A nontrivial boundary value problem,''
  Phys.\ Rev.\ D {\bf 87}, 104001 (2013)
  [arXiv:1208.5902 [hep-th]].
%\cite{Bai:2012cx}
\bibitem{Bai:2012cx} 
  N.~Bai, Y.~H.~Gao, B.~G.~Qi and X.~B.~Xu,
  ``Holographic insulator/superconductor phase transition in Born-Infeld electrodynamics,''
  arXiv:1212.2721 [hep-th].
%\cite{Zhao:2012kp}
\bibitem{Zhao:2012kp} 
  Z.~Zhao, Q.~Pan and J.~Jing,
  ``Holographic insulator/superconductor phase transition with Weyl corrections,''
  Phys.\ Lett.\ B {\bf 719}, 440 (2013)
  [arXiv:1212.3062].
%\cite{Gangopadhyay:2013zph}
\bibitem{Gangopadhyay:2013zph} 
  S.~Gangopadhyay,
  ``Analytic study of properties of holographic superconductors away from the probe limit,''
  Phys.\ Lett.\ B {\bf 724}, 176 (2013)
  [arXiv:1302.1288 [hep-th]].
%\cite{Momeni:2013waa}
\bibitem{Momeni:2013waa}
  D.~Momeni, M.~Raza, M.~R.~Setare and R.~Myrzakulov,
  ``Analytical Holographic Superconductor with Backreaction Using AdS3/CFT2,''
  Int.\ J.\ Theor.\ Phys.\  {\bf 52} (2013) 2773
  [arXiv:1305.5163 [physics.gen-ph]].
%\cite{Li:2013glh}
\bibitem{Li:2013glh} 
  H.~F.~Li,
  ``Further studies on holographic insulator/superconductor phase transitions from Sturm-Liouville eigenvalue problems,''
  JHEP {\bf 1307}, 135 (2013)
  [arXiv:1306.3071 [hep-th]].
%\cite{Lala:2014jca}
\bibitem{Lala:2014jca} 
  A.~Lala,
  ``Magnetic response of holographic Lifshitz superconductors:Vortex and Droplet solutions,''
  Phys.\ Lett.\ B {\bf 735}, 396 (2014)
  [arXiv:1404.2774 [hep-th]].

%%%%%%%%%%%%%%%%%%%%%%
%s-wave calculations
%\cite{Momeni:2011ca}
\bibitem{Momeni:2011ca} 
  D.~Momeni and M.~R.~Setare,
  ``A note on holographic superconductors with Weyl Corrections,''
  Mod.\ Phys.\ Lett.\ A {\bf 26}, 2889 (2011)
  [arXiv:1106.0431 [physics.gen-ph]].

%%%%%%%%%%%%%%%%%%%%%%%%
%Stuckelberg
\bibitem{Ma:2011zze} 
  D.~Z.~Ma, Y.~Cao and J.~P.~Wu,
  %``The St\'{u}ckelberg Holographic Superconductors with Weyl corrections,''
  Phys.\ Lett.\ B {\bf 704}, 604 (2011)
  [arXiv:1201.2486 [hep-th]].  
%\cite{Momeni:2012uc}
\bibitem{Momeni:2012uc} 
  D.~Momeni, M.~R.~Setare and R.~Myrzakulov,
  %``Condensation of the scalar field with Stuckelberg and Weyl Corrections in the background of a planar AdS-Schwarzschild black hole,''
  Int.\ J.\ Mod.\ Phys.\ A {\bf 27}, 1250128 (2012)
  [arXiv:1209.3104 [physics.gen-ph]].
%%%%%%%%%%%%%%%%%%%%%%%%%%%%
%p-wave-Weyl

%\cite{Momeni:2012ab}
\bibitem{Momeni:2012ab} 
  D.~Momeni, N.~Majd and R.~Myrzakulov,
  %``p-wave holographic superconductors with Weyl corrections,''
  Europhys.\ Lett.\  {\bf 97}, 61001 (2012)
  [arXiv:1204.1246 [hep-th]].

%\cite{Momeni:2013fma}
\bibitem{Momeni:2013fma} 
  D.~Momeni, R.~Myrzakulov and M.~Raza,
  ``Holographic superconductors with Weyl Corrections via gauge/gravity duality,''
  Int.\ J.\ Mod.\ Phys.\ A {\bf 28}, 1350096 (2013)
  [arXiv:1307.8348 [hep-th]].

\bibitem{Gorski:1983pp} 
  A.~Gorski,
  %``Strong Coupling Expansion For Classical Yang-mills Theory,''
  J.\ Phys.\ A {\bf 16}, 849 (1983).


\bibitem{Dufaux:2004qs} 
  J.~F.~Dufaux, J.~E.~Lidsey, R.~Maartens and M.~Sami,
  %``Cosmological perturbations from brane inflation with a Gauss-Bonnet term,''
  Phys.\ Rev.\ D {\bf 70}, 083525 (2004)
  [hep-th/0404161].

%\cite{Tsujikawa:2004dm}
\bibitem{Tsujikawa:2004dm} 
  S.~Tsujikawa, M.~Sami and R.~Maartens,
  %``Observational constraints on braneworld inflation: The Effect of a Gauss-Bonnet term,''
  Phys.\ Rev.\ D {\bf 70}, 063525 (2004)
  [astro-ph/0406078].


\end{thebibliography}
\end{document}